\pdfoutput=1
\documentclass[12pt,a4paper]{article}
\usepackage{ifthen} % for conditional statements
\newboolean{pdflatex}
\setboolean{pdflatex}{true} % False for eps figures 

\newboolean{articletitles}
\setboolean{articletitles}{true} % False removes titles in references

\newboolean{uprightparticles}
\setboolean{uprightparticles}{false} %True for upright particle symbols

\def\paperauthors{LHCb collaboration} % Leave as is for PAPER, CONF and FIGURE
\def\paperasciititle{Measurement of the Omegab- baryon lifetime} % Set ASCII title here !! MAKE sure it's only ASCII characters !! 
\def\papertitle{Measurement of the $\Omegab$ baryon lifetime} % Latex formatted title
\def\paperkeywords{{High Energy Physics}, {LHCb}} % Comma separated list
\def\papercopyright{\the\year\ CERN for the benefit of the LHCb collaboration} % new since 9/Apr/2018
\def\paperlicence{CC BY 4.0 licence}
\def\paperlicenceurl{https://creativecommons.org/licenses/by/4.0/}

\newif\ifEnableSectionTOCLinks
\EnableSectionTOCLinksfalse % deactivated

\usepackage[top=1in, bottom=1.25in, left=1in, right=1in]{geometry}

\usepackage{microtype}
\usepackage{lineno}  % for line numbering during review
\usepackage{xspace} % To avoid problems with missing or double spaces after
\usepackage{caption} %these three command get the figure and table captions automatically small

\usepackage{graphicx}  % to include figures (can also use other packages)
\usepackage{color}
\usepackage{colortbl}
\graphicspath{{./figs/}} % Make Latex search fig subdir for figures
\usepackage{amsmath} % Adds a large collection of math symbols
\usepackage{amssymb}
\usepackage{amsfonts}
\usepackage{upgreek} % Adds in support for greek letters in roman typeset

\usepackage[normalem]{ulem} % for \sout strikeout in template tables

\newcommand*\patchAmsMathEnvironmentForLineno[1]{%
\expandafter\let\csname old#1\expandafter\endcsname\csname #1\endcsname
\expandafter\let\csname oldend#1\expandafter\endcsname\csname
end#1\endcsname
 \renewenvironment{#1}%
   {\linenomath\csname old#1\endcsname}%
   {\csname oldend#1\endcsname\endlinenomath}%
}
\newcommand*\patchBothAmsMathEnvironmentsForLineno[1]{%
  \patchAmsMathEnvironmentForLineno{#1}%
  \patchAmsMathEnvironmentForLineno{#1*}%
}
\AtBeginDocument{%
\patchBothAmsMathEnvironmentsForLineno{equation}%
\patchBothAmsMathEnvironmentsForLineno{align}%
\patchBothAmsMathEnvironmentsForLineno{flalign}%
\patchBothAmsMathEnvironmentsForLineno{alignat}%
\patchBothAmsMathEnvironmentsForLineno{gather}%
\patchBothAmsMathEnvironmentsForLineno{multline}%
\patchBothAmsMathEnvironmentsForLineno{eqnarray}%
}

\usepackage[pdftex,
            pdfauthor={\paperauthors},
            pdftitle={\paperasciititle},
            pdfkeywords={\paperkeywords}]{hyperref}
\usepackage{hyperxmp}
\hypersetup{
    pdfcopyright={Copyright (C) \papercopyright},
    pdflicenseurl={\paperlicenceurl}
}
\usepackage[colorinlistoftodos,textsize=scriptsize]{todonotes}

\usepackage[bottom,flushmargin,hang,multiple]{footmisc}

\usepackage[all]{hypcap} % Internal hyperlinks to floats.

\usepackage{xspace} 
\usepackage{upgreek}

\def\lhcb   {\mbox{LHCb}\xspace}

\def\MagUp {\mbox{\em Mag\kern -0.05em Up}\xspace}

\ifthenelse{\boolean{uprightparticles}}%
{

 \def\Pmu         {\ensuremath{\upmu}\xspace}                 
 \def\Pnu         {\ensuremath{\upnu}\xspace}                 
                  
 \def\Ppi         {\ensuremath{\uppi}\xspace}                 
                  
 \def\Prho        {\ensuremath{\uprho}\xspace}

 \def\Ppsi        {\ensuremath{\uppsi}\xspace}

 \def\PDelta      {\ensuremath{\Delta}\xspace}                 
 \def\PXi         {\ensuremath{\Xi}\xspace}                 
 \def\PLambda     {\ensuremath{\Lambda}\xspace}                 
 \def\PSigma      {\ensuremath{\Sigma}\xspace}                 
 \def\POmega      {\ensuremath{\Omega}\xspace}                 
 \def\PUpsilon    {\ensuremath{\Upsilon}\xspace}
 \let\oldPi\Pi
 \def\PPi         {\ensuremath{\oldPi}\xspace}

 \def\PB      {\ensuremath{\mathrm{B}}\xspace}                 
                  
 \def\PD      {\ensuremath{\mathrm{D}}\xspace}

 \def\PJ      {\ensuremath{\mathrm{J}}\xspace}                 
 \def\PK      {\ensuremath{\mathrm{K}}\xspace}

 \def\PW      {\ensuremath{\mathrm{W}}\xspace}

 \def\Pb      {\ensuremath{\mathrm{b}}\xspace}                 
 \def\Pc      {\ensuremath{\mathrm{c}}\xspace}

 \def\Pi      {\ensuremath{\mathrm{i}}\xspace}

 \def\Pp      {\ensuremath{\mathrm{p}}\xspace}

 \def\Ps      {\ensuremath{\mathrm{s}}\xspace}

 \def\thebaroffset{0.0em}
}
{

 \def\Pmu         {\ensuremath{\mu}\xspace}                 
 \def\Pnu         {\ensuremath{\nu}\xspace}                 
                  
 \def\Ppi         {\ensuremath{\pi}\xspace}                 
                  
 \def\Prho        {\ensuremath{\rho}\xspace}

 \def\Ppsi        {\ensuremath{\psi}\xspace}                 
                  
 \mathchardef\PDelta="7101
 \mathchardef\PXi="7104
 \mathchardef\PLambda="7103
 \mathchardef\PSigma="7106
 \mathchardef\POmega="710A
 \mathchardef\PUpsilon="7107
 \mathchardef\PPi="7105
                  
 \def\PB      {\ensuremath{B}\xspace}                 
                  
 \def\PD      {\ensuremath{D}\xspace}

 \def\PJ      {\ensuremath{J}\xspace}                 
 \def\PK      {\ensuremath{K}\xspace}

 \def\PW      {\ensuremath{W}\xspace}

 \def\Pb      {\ensuremath{b}\xspace}                 
 \def\Pc      {\ensuremath{c}\xspace}

 \def\Pi      {\ensuremath{i}\xspace}

 \def\Pp      {\ensuremath{p}\xspace}

 \def\Ps      {\ensuremath{s}\xspace}

 \def\thebaroffset{0.18em}
}
\newcommand{\offsetoverline}[2][\thebaroffset]{\kern #1\overline{\kern -#1 #2}}%

\makeatletter
\ifcase \@ptsize \relax% 10pt
  \newcommand{\miniscule}{\@setfontsize\miniscule{4}{5}}% \tiny: 5/6
\or% 11pt
  \newcommand{\miniscule}{\@setfontsize\miniscule{5}{6}}% \tiny: 6/7
\or% 12pt
  \newcommand{\miniscule}{\@setfontsize\miniscule{5}{6}}% \tiny: 6/7
\fi
\makeatother

\DeclareRobustCommand{\optbar}[1]{\shortstack{{\miniscule (\rule[.5ex]{1.25em}{.18mm})}
  \\ [-.7ex] $#1$}}

\def\mup        {{\ensuremath{\Pmu^+}}\xspace}
\def\mun        {{\ensuremath{\Pmu^-}}\xspace} % muon negative (\mum is taken)

\def\ellm       {{\ensuremath{\ell^-}}\xspace}

\def\neub       {{\ensuremath{\overline{\Pnu}}}\xspace}

\def\neulb      {{\ensuremath{\neub_\ell}}\xspace}
\def\Wm     {{\ensuremath{\PW^-}}\xspace}

\def\squark    {{\ensuremath{\Ps}}\xspace}

\def\cquark    {{\ensuremath{\Pc}}\xspace}

\def\bquark    {{\ensuremath{\Pb}}\xspace}
\def\bquarkbar {{\ensuremath{\overline \bquark}}\xspace}
\def\bbbar     {{\ensuremath{\bquark\bquarkbar}}\xspace}

\def\pion   {{\ensuremath{\Ppi}}\xspace}

\def\pip    {{\ensuremath{\pion^+}}\xspace}
\def\pim    {{\ensuremath{\pion^-}}\xspace}

\def\rhomeson {{\ensuremath{\Prho}}\xspace}

\def\rhom     {{\ensuremath{\rhomeson^-}}\xspace}

\def\kaon    {{\ensuremath{\PK}}\xspace}
\def\Kbar    {{\ensuremath{\offsetoverline{\PK}}}\xspace}

\def\KorKbar {\kern \thebaroffset\optbar{\kern -\thebaroffset \PK}{}\xspace}

\def\Km      {{\ensuremath{\kaon^-}}\xspace}

\def\Kstarzb {{\ensuremath{\Kbar{}^{*0}}}\xspace}

\def\Xicprime    {{\ensuremath{\Xires^{'0}_\cquark}}\xspace}
\def\OmegacStar      {{\ensuremath{\Omegares(2770)^{0}_\cquark}}\xspace}

\def\D       {{\ensuremath{\PD}}\xspace}

\def\DorDbar {\kern \thebaroffset\optbar{\kern -\thebaroffset \PD}\xspace}
\def\Dz      {{\ensuremath{\D^0}}\xspace}

\def\Dp      {{\ensuremath{\D^+}}\xspace}
\def\Dm      {{\ensuremath{\D^-}}\xspace}

\def\DpDm    {\ensuremath{\Dp {\kern -0.16em \Dm}}\xspace}

\def\Dstarp  {{\ensuremath{\D^{*+}}}\xspace}

\def\B       {{\ensuremath{\PB}}\xspace}

\def\BorBbar {\kern \thebaroffset\optbar{\kern -\thebaroffset \PB}\xspace}

\def\Bd      {{\ensuremath{\B^0}}\xspace}

\def\BdorBdbar {\kern \thebaroffset\optbar{\kern -\thebaroffset \Bd}\xspace}

\def\Bs      {{\ensuremath{\B^0_\squark}}\xspace}

\def\BsorBsbar {\kern \thebaroffset\optbar{\kern -\thebaroffset \Bs}\xspace}

\def\jpsi     {{\ensuremath{{\PJ\mskip -3mu/\mskip -2mu\Ppsi}}}\xspace}

\def\Y#1S{\ensuremath{\PUpsilon{(#1S)}}\xspace}

\def\proton      {{\ensuremath{\Pp}}\xspace}

\def\Lz          {{\ensuremath{\PLambda}}\xspace}
\def\LzStar          {{\ensuremath{\PLambda(1520)}}\xspace}

\def\LorLbar     {\kern \thebaroffset\optbar{\kern -\thebaroffset \PLambda}\xspace}

\def\Xires       {{\ensuremath{\PXi}}\xspace}

\def\Xim         {{\ensuremath{\Xires^-}}\xspace}

\def\Omegares    {{\ensuremath{\POmega}}\xspace}

\def\Omegam      {{\ensuremath{\Omegares^-}}\xspace}

\def\Lc          {{\ensuremath{\Lz^+_\cquark}}\xspace}

\def\Xic         {{\ensuremath{\Xires_\cquark}}\xspace}
\def\Xicz        {{\ensuremath{\Xires^0_\cquark}}\xspace}

\def\Omegac      {{\ensuremath{\Omegares^0_\cquark}}\xspace}
\def\Omegacstar  {{\ensuremath{\Omegares_\cquark(2770)^{0}}}\xspace}
\def\OmegacStar  {{\ensuremath{\Omegares_\cquark(2770)^{0}}}\xspace}

\def\Lb           {{\ensuremath{\Lz^0_\bquark}}\xspace}

\def\Xib          {{\ensuremath{\Xires_\bquark}}\xspace}
\def\Xibz         {{\ensuremath{\Xires^0_\bquark}}\xspace}
\def\Xibm         {{\ensuremath{\Xires^-_\bquark}}\xspace}

\def\Omegab       {{\ensuremath{\Omegares^-_\bquark}}\xspace}

\newcommand{\decay}[2]{\ensuremath{#1\!\to #2}\xspace} 

\def\to                 {\ensuremath{\rightarrow}\xspace}

\def\eff   {{\ensuremath{\varepsilon}}\xspace}

\def\AT#1     {\ensuremath{A_{\mathrm{T}}^{#1}}\xspace}           % 2

\def\C#1      {\ensuremath{\mathcal{C}_{#1}}\xspace}                       % 9
\def\Cp#1     {\ensuremath{\mathcal{C}_{#1}^{'}}\xspace}                    % 7
\def\Ceff#1   {\ensuremath{\mathcal{C}_{#1}^{\mathrm{(eff)}}}\xspace}        % 9  
\def\Cpeff#1  {\ensuremath{\mathcal{C}_{#1}^{'\mathrm{(eff)}}}\xspace}       % 7
\def\Ope#1    {\ensuremath{\mathcal{O}_{#1}}\xspace}                       % 2
\def\Opep#1   {\ensuremath{\mathcal{O}_{#1}^{'}}\xspace}                    % 7

\newcommand{\nospaceunit}[1]{\ensuremath{\text{#1}}}       
\newcommand{\aunit}[1]{\ensuremath{\text{\,#1}}}       
\newcommand{\tev}{\aunit{Te\kern -0.1em V}\xspace}
\newcommand{\gev}{\aunit{Ge\kern -0.1em V}\xspace}
\newcommand{\mev}{\aunit{Me\kern -0.1em V}\xspace}
\newcommand{\kev}{\aunit{ke\kern -0.1em V}\xspace}
\newcommand{\ev}{\aunit{e\kern -0.1em V}\xspace}
 
\newcommand{\mevc}{\ensuremath{\aunit{Me\kern -0.1em V\!/}c}\xspace}
\newcommand{\gevc}{\ensuremath{\aunit{Ge\kern -0.1em V\!/}c}\xspace}
\newcommand{\mevcc}{\ensuremath{\aunit{Me\kern -0.1em V\!/}c^2}\xspace}
\newcommand{\gevcc}{\ensuremath{\aunit{Ge\kern -0.1em V\!/}c^2}\xspace}
\def\mum  {\ensuremath{\,\upmu\nospaceunit{m}}\xspace}

\def\fb   {\ensuremath{\aunit{fb}}\xspace}
\def\invfb   {\ensuremath{\fb^{-1}}\xspace}

\def\ps   {\ensuremath{\aunit{ps}}\xspace}

\newcommand{\chisq}{\ensuremath{\chi^2}\xspace}

\newcommand{\chisqip}{\ensuremath{\chi^2_{\text{IP}}}\xspace}

\def\gsim{{~\raise.15em\hbox{$>$}\kern-.85em
          \lower.35em\hbox{$\sim$}~}\xspace}
\def\lsim{{~\raise.15em\hbox{$<$}\kern-.85em
          \lower.35em\hbox{$\sim$}~}\xspace}

\def\sPlot{\mbox{\em sPlot}\xspace}

\def\pt         {\ensuremath{p_{\mathrm{T}}}\xspace}

\def\ptot       {\ensuremath{p}\xspace}

\def\evtgen     {\mbox{\textsc{EvtGen}}\xspace}

\def\geant      {\mbox{\textsc{Geant4}}\xspace}

\def\photos     {\mbox{\textsc{Photos}}\xspace}

\def\pythia     {\mbox{\textsc{Pythia}}\xspace}

\def\tell1  {TELL1\xspace}
\def\ukl1   {UKL1\xspace}

\newcommand{\lhcborcid}[1]{\href{https://orcid.org/#1}{\hspace*{0.1em}\raisebox{-0.45ex}{\includegraphics[width=1em]{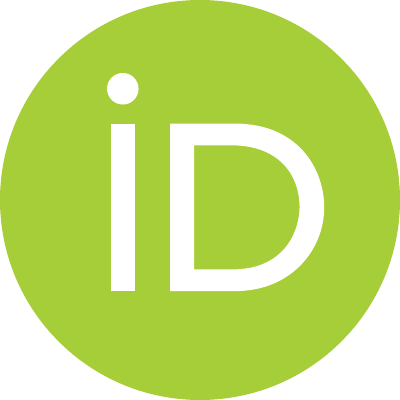}}}}

\hypersetup{
  colorlinks   = true, %Colours links instead of ugly boxes
  urlcolor     = blue, %Colour for external hyperlinks
  linkcolor    = blue, %Colour of internal links
  citecolor    = red   %Colour of citations
}

\ifEnableSectionTOCLinks
    \usepackage[explicit]{titlesec} % to change headings
    
    \let\oldcontentsline\contentsline
    \renewcommand

    \titleformat{\section}{\normalfont\Large\bf}{\hyperlink{tocsection.\thesection}{{\thesection} \parbox[t]{\dimexpr\textwidth-1pc}{#1}}}{1pc}{}

    \titleformat{\subsection}{\normalfont\bf}{\hyperlink{tocsubsection.\thesubsection}{{\thesubsection} \parbox[t]{\dimexpr\textwidth-1pc}{#1}}}{1pc}{}

    \titleformat{name=\section,numberless}[display]{}{}{0pt}{\normalfont\Huge\bfseries #1}
\fi

\usepackage{cite} % Allows for ranges in citations
\usepackage{LHCb/mciteplus}
\makeatletter
\g@addto@macro\bfseries{\boldmath}
\makeatother
\usepackage{longtable} % only for template; not usually to be used in PAPERs

\begin{document}

%%%%%%%%%%%%%%%%%%%%%%%%%
%%%%% Title     %%%%%%%%%
%%%%%%%%%%%%%%%%%%%%%%%%%
\renewcommand{\thefootnote}{\fnsymbol{footnote}}
\setcounter{footnote}{1}

% %%%%%%% CHOOSE TITLE PAGE--------
%\onecolumn
%\input{title-LHCb-INT}
%\input{title-LHCb-ANA}
%\input{title-LHCb-CONF}
%\input{title-LHCb-FIGURE}
% ===============================================================================
% Purpose: LHCb-PAPER journal paper title page template
% Author: 
% Created on: 2010-09-25
% ===============================================================================

%%%%%%%%%%%%%%%%%%%%%%%%%
%%%%%  TITLE PAGE  %%%%%%
%%%%%%%%%%%%%%%%%%%%%%%%%
\begin{titlepage}
\pagenumbering{roman}

% Header ---------------------------------------------------
\vspace*{-1.5cm}
\centerline{\large EUROPEAN ORGANIZATION FOR NUCLEAR RESEARCH (CERN)}
\vspace*{1.5cm}
\noindent
\begin{tabular*}{\linewidth}{lc@{\extracolsep{\fill}}r@{\extracolsep{0pt}}}
\ifthenelse{\boolean{pdflatex}}% Logo format choice
{\vspace*{-1.5cm}\mbox{\!\!\!\includegraphics[width=.14\textwidth]{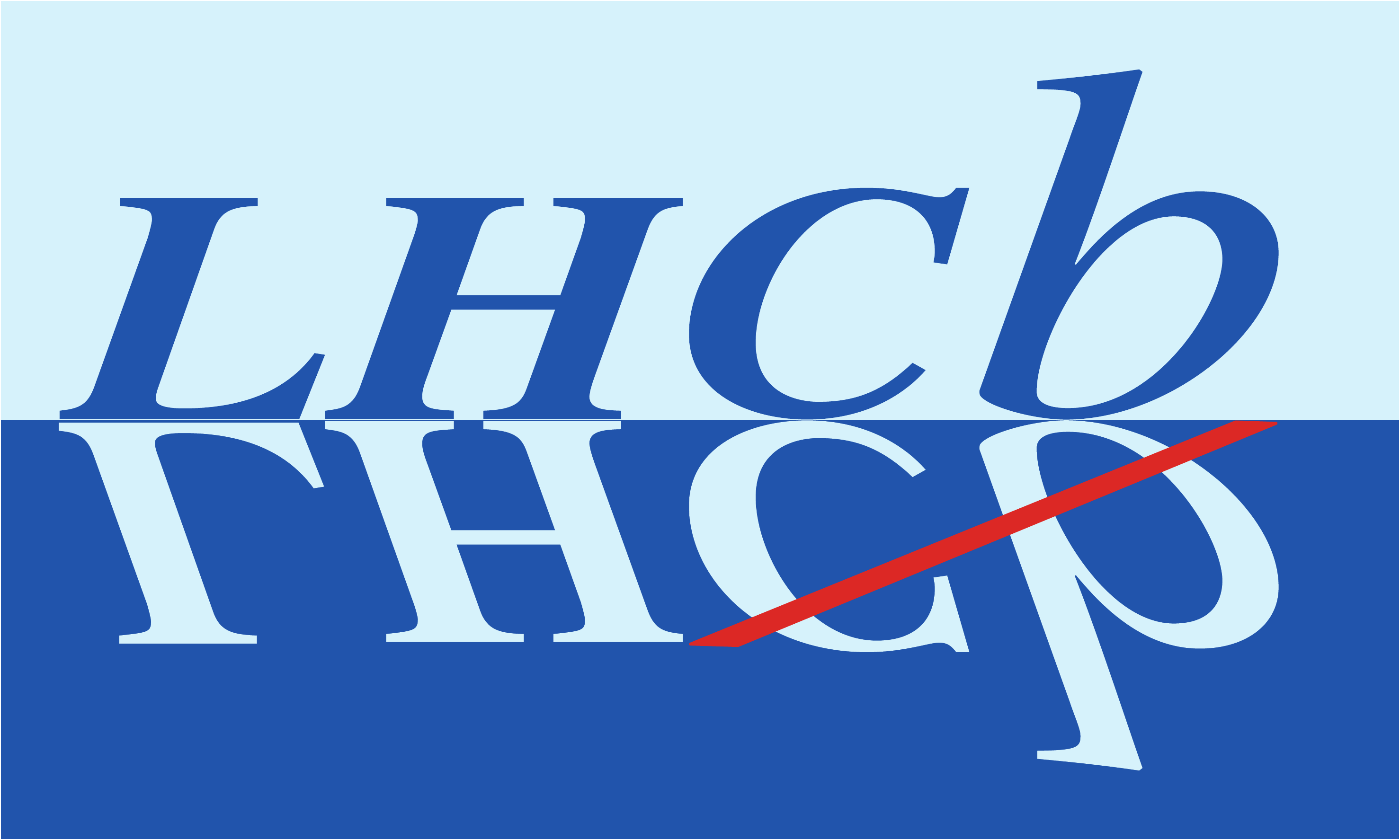}} & &}%
{\vspace*{-1.2cm}\mbox{\!\!\!\includegraphics[width=.12\textwidth]{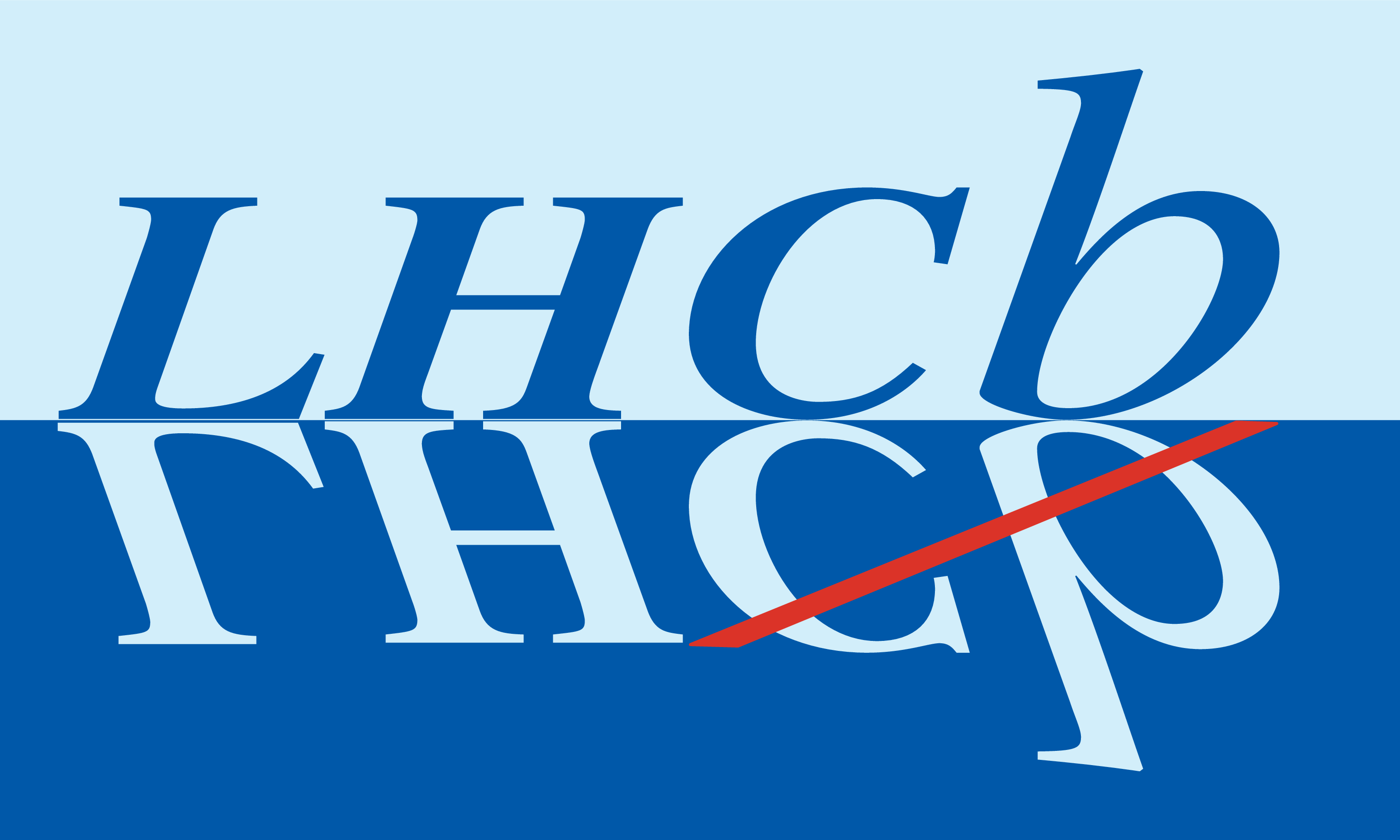}} & &}%
\\
 & & CERN-EP-2026-246 \\  % ID 
 & & LHCb-PAPER-2026-034 \\  % ID 
 & & September 21, 2026 \\ % Date - Can also hardwire e.g.: 23 March 2010
 & & \\
% not in paper \hline
\end{tabular*}

\vspace*{4.0cm}

% Title --------------------------------------------------
{\normalfont\bfseries\boldmath\huge
\begin{center}
% DO NOT EDIT HERE. Instead edit macro in main.tex to keep metadata correct
  \papertitle 
\end{center}
}

\vspace*{2.0cm}

% Authors -------------------------------------------------
\begin{center}
%In the footnote, replace 'paper' by 'Letter' in case of submission to PRL or PLB 
% Edit macro in main.tex to keep metadata correct
\paperauthors\footnote{Authors are listed at the end of this paper.}
\end{center}

\vspace{\fill}

% Abstract -----------------------------------------------
\begin{abstract}
  \noindent
The lifetime ratio ${r_{\tau}\equiv\tau_{\Omegab}/\tau_{\Xibm}}$ between the ${{\Omegab}}$ and ${{\Xibm}}$ baryons is measured using a sample of $pp$ collision data corresponding to an integrated luminosity of 6~fb$^{-1}$ and collected by the LHCb experiment during LHC Run 2 (2015--2018). The ratio $r_{\tau}$ is measured in two sets of decays modes, \mbox{${ {(\Omegab,\Xibm)\to(\Omegac\pi^-,\Xicz\pi^-)}}$} and \mbox{${{(\Omegab,\Xibm)\to(J/\psi\Omegam, \jpsi\Xim)}}$}, with \mbox{${{(\Omegac,\Xicz)\to pK^-K^-\pi^+}}$, ${{(\Omegam,\Xim)\to(\Lz K^-,\Lz\pi^-)}}$, ${{\Lz\to p\pi^-}}$} and \mbox{$\jpsi\to\mup\mun$}. The measured $r_{\tau}$ values are averaged and combined with Run 1 (2011--2012) measurements in the same decay modes to obtain ${r_{\tau} = 1.109\pm0.055\pm0.010}$. Multiplying by the known ${{\Xibm}}$ lifetime results in the ${{\Omegab}}$ lifetime ${\tau_{\Omegab} = 1.751\pm0.089\pm0.022~{\rm ps}}$, where the uncertainties are statistical and systematic. This measurement improves on the precision of the $\Omegab$ lifetime by about a factor of two over the previous world average. The value of $r_{\tau}$ is in agreement with the most recent theoretical predictions from the heavy quark expansion framework.
  
\end{abstract}

\vspace*{2.0cm}

\begin{center}
  Submitted to
  Phys.~Rev.~D 
\end{center}

\vspace{\fill}

{\footnotesize 
% Edit macro in main.tex to keep metadata correct
\centerline{\copyright~\papercopyright. \href{\paperlicenceurl}{\paperlicence}.}}
\vspace*{2mm}

\end{titlepage}

%%%%%%%%%%%%%%%%%%%%%%%%%%%%%%%%
%%%%%  EOD OF TITLE PAGE  %%%%%%
%%%%%%%%%%%%%%%%%%%%%%%%%%%%%%%%

%  empty page follows the title page ----
\newpage
\setcounter{page}{2}
\mbox{~}
%\newpage
%
%% Author List ----------------------------
%%  You need to get a new author list!
%\input{LHCb_authorlist.tex}
%
%The author list for journal publications is provided by the Membership Committee shortly after 'approval to go to paper' has been given.
%%It will be made available on the page
%%\verb!http://www.physik.uzh.ch/~strauman/forMemCo/LHCb-PAPER-XXXX-XXX/! .
%It will be sent to you by email shortly after a paper number has beens assigned.
%The author list should be included already at first circulation, 
%to allow new members of the collaboration to verify whether they have been included correctly.
%Occasionally a misspelled name is corrected or associated institutions become full members.
%In that case, a new author list will be sent to you.
%In case line numbering doesn't work well after including the authorlist, try moving the \verb!\bigskip! after the last author to a separate line.
%
%
%The authorship for Conference Reports should be ``The LHCb
%  collaboration'', with a footnote giving the name(s) of the contact
%  author(s), but without the full list of collaboration names.

%\twocolumn
% %%%%%%%%%%%%% ---------

\renewcommand{\thefootnote}{\arabic{footnote}}
\setcounter{footnote}{0}

%%%%%%%%%%%%%%%%%%%%%%%%%%%%%%%%
%%%%%  Table of Content   %%%%%%
%%%%%%%%%%%%%%%%%%%%%%%%%%%%%%%%
%%%% Uncomment if desired
%\tableofcontents

\cleardoublepage

%%%%%%%%%%%%%%%%%%%%%%%%%
%%%%% Main text %%%%%%%%%
%%%%%%%%%%%%%%%%%%%%%%%%%

\pagestyle{plain} % restore page numbers for the main text
\setcounter{page}{1}
\pagenumbering{arabic}

%% Uncomment during review phase. 
%% Comment before a final submission.
%\linenumbers

%% This is the main body
%% It is useful to have a single file so comments are not missed in overleaf.
\section{Introduction}
\label{sec:Introduction}

The heavy quark expansion
(HQE)~\cite{Lenz:2014jha} is a theoretical framework that predicts the inclusive decay rates of beauty hadrons through an expansion in powers of the strong coupling constant, $\alpha_s$, and $\Lambda_{\rm QCD}/m_b$. Here, $\Lambda_{\rm QCD}$ is the energy scale below which the strong-interaction coupling becomes large, ${\mathcal{O}}(100\mev)$, and $m_b$ is the $b$-quark mass. When combined with precision measurements of heavy-quark decays, the HQE framework can be used to calculate $b$-hadron parameters required for the determination of Cabibbo-Kobayashi-Maskawa~\cite{Cabibbo:1963yz,Kobayashi:1973fv} matrix elements, which in turn provide constraints on physics beyond the Standard Model.

Measurements of the lifetimes ($\tau$) of hadrons containing heavy quarks provide a stringent test of the HQE framework through the calculation of the total decay width, \mbox{$\Gamma=\hbar/\tau$}. At leading order within the HQE framework, all weakly-decaying hadrons containing a single $b$ quark have decay widths equal to that of the $b$ quark, and therefore equal lifetimes. Nonperturbative effects, described by the kinetic, Darwin and chromomagnetic operators~\cite{Gratrex,Dassinger_2007,SumRules,HigherOrderCorr}, coherently shift the decay widths of all $b$ hadrons from that of the free $b$ quark. A second set of higher order corrections is due to the interactions of the $b$ quark with the light valence quark(s), and these effects lead to differences in the total decay width of various $b$ hadrons. The beauty baryons, $\Lb(bud)$, $\Xibz(bus)$, $\Xibm(bds)$ and $\Omegab(bss)$ provide a fertile testing ground for the HQE predictions due to their different pairs of valence quarks. The latest measurements of the lifetime ratios $\tau_{\Xibm}/\tau_{\Lb}$~\cite{LHCb-PAPER-2024-010} and $\tau_{\Xibz}/\tau_{\Lb}$~\cite{LHCb-PAPER-2025-023} have reached 1\% precision, and their values are in good agreement with recent calculations from the HQE framework~\cite{lenz2026predictionsbbaryonlifetimesnnloqcd}.

A precision test of HQE predictions with the $\Omegab$ baryon is limited by the large experimental uncertainty on its lifetime.  Collecting large samples of $\Omegab$ baryons at the LHC is challenging due to the low production rate compared to other beauty baryons. Nevertheless, this test is important, as the $\Omegab$ baryon has unique features compared to the $\Lb$, $\Xibz$ and $\Xibm$ baryons, most notably the presence of two valence $s$ quarks in a spin 1 state, compared to the spin 0 state of the other ground-state $b$ baryons. The spin 1 contribution probes the chromomagnetic operator in the HQE framework, which is absent for the other beauty baryons~\cite{Gratrex}. Moreover, interest in improved measurements is reinforced by the recent measurements of the large $\Omegac$ lifetime~\cite{LHCb-PAPER-2018-028,LHCb-PAPER-2021-021,LHCb-PAPER-2025-013}. The current world average value of the $\Omegab$ lifetime, $\tau_{\Omegab}=1.64\pm0.16$~ps~\cite{PDG2026}, is obtained from a measurement of the lifetime ratio $r_{\tau}\equiv\tau_{\Omegab}/\tau_{\Xibm}$ by the LHCb collaboration~\cite{LHCb-PAPER-2016-008} using $\Omegab\to\Omegac\pim$ decays, and a set of measurements of the absolute lifetime by the CDF~\cite{CDF:2014mon} and LHCb~\cite{LHCb-PAPER-2014-010} collaborations using the $\Omegab\to\jpsi\Omegam$ decay. 

In this paper, a new measurement of $r_{\tau}$ is reported. The analysis uses a $pp$ collision data sample collected between 2015 and 2018 (Run\,2) by the LHCb experiment at a center-of-mass energy of $\sqrt{s}=13\tev$, corresponding to an integrated luminosity of 6\invfb. Two sets of decay modes are analyzed:
$(\Omegab,\Xibm)\to(\Omegac\pim,\Xicz\pim)$ and $(\Omegab,\Xibm)\to(\jpsi\Omegam, \jpsi\Xim)$, with $(\Omegac,\Xicz)\to p\Km\Km\pip$, $(\Omegam,\Xim)\to(\Lz\Km,\Lz\pim)$, $\Lz\to p\pim$ and $\jpsi\to\mup\mun$. Inclusion of charge-conjugate processes is implied throughout this paper. The $\Omegac\pim$ and $\jpsi\Omegam$ final states provide the largest yields of fully reconstructed $\Omegab$ decays in LHCb. The similarity of the $\Omegab$ and $\Xibm$ final states leads to a reduction in the systematic uncertainty in $r_{\tau}$. The integrated luminosity and $\bbbar$ production cross section are each about a factor of two larger than those of the previous measurements~\cite{LHCb-PAPER-2014-010,LHCb-PAPER-2016-008}. 

The quantity that is measured experimentally is the efficiency-corrected ratio of signal yields as a function of decay time, $t$,
\begin{align}
    R(t) \equiv\frac{N[\Omegab\to f_s](t)}{N[\Xibm\to f_n](t)} \cdot \frac{\eff[\Xibm\to f_n](t)}{\eff[\Omegab\to f_s](t)}=R_0\exp{(\lambda t)},
\label{eq:RT}
\end{align}
\noindent where $R_0$ is an overall normalization factor,  $f_s$ and $f_n$ represent the final states in the signal and normalization modes,
$(f_s,f_n)=(\Omegac\pim,\Xicz\pim)$ and $(\jpsi\Omegam, \jpsi\Xim)$, and
$N$ and $\eff$ represent the observed signal yields and efficiencies of those decays. The parameter $\lambda$ is related to the lifetimes of the $\Omegab$ and $\Xibm$ baryons through
\begin{align}
    \lambda\equiv\frac{1}{\tau_{\Xibm}}-\frac{1}{\tau_{\Omegab}}.
\end{align}
\noindent The ratio of lifetimes is then
\begin{align}
   r_{\tau}=\frac{\tau_{\Omegab}}{\tau_{\Xibm}}=\frac{1}{1-\lambda\tau_{\Xibm}}.
\label{eq:rtau} 
\end{align}
\noindent Based on the known values $\tau_{\Omegab}=1.64\pm0.16$~ps~\cite{PDG2026} and $\tau_{\Xibm}=1.579\pm0.023\ps$~\cite{LHCb-PAPER-2024-010,PDG2026},
the $\lambda$ parameter is expected to be small, with magnitude of order $0.06\ps^{-1}$. Consequently, the uncertainty in the $\Xibm$ baryon lifetime is a small contribution to the overall uncertainty in~$r_{\tau}$.

\section{Detector and simulation}
The \lhcb detector~\cite{LHCb-DP-2008-001,LHCb-DP-2014-002} is a single-arm forward
spectrometer covering the \mbox{pseudorapidity} range $2<\eta <5$, designed for the study of particles containing \bquark or \cquark
quarks. The detector used for this analysis includes a high-precision tracking system
consisting of a silicon-strip vertex detector (VELO) surrounding the $pp$
interaction region~\cite{LHCb-DP-2014-001}, a large-area silicon-strip detector (TT) located
upstream of a dipole magnet with a bending power of about
$4{\mathrm{\,T\,m}}$, and three stations of silicon-strip detectors and straw
drift tubes placed downstream of the magnet (T-stations).
The tracking system provides a measurement of the momentum, \ptot, of charged particles with
a relative uncertainty that varies from 0.5\% at low momentum to 1.0\% at 200\gevc. The polarity of the LHCb magnet is alternated regularly throughout each period of data taking. The minimum distance of a track to a primary $pp$ collision vertex (PV), the impact parameter~(IP), 
is measured with a resolution of approximately $\sigma_{\rm IP}=(15+29/\pt)\mum$,
where \pt is the component of the momentum transverse to the beam, in\,\gevc.
Different types of charged hadrons are distinguished using information
from two ring-imaging Cherenkov detectors~\cite{LHCb-DP-2012-003}. 
Photons, electrons and hadrons are identified by a calorimeter system consisting of
scintillating-pad and preshower detectors, an electromagnetic
and a hadronic calorimeter. Muons are identified by a
system composed of alternating layers of iron and multiwire
proportional chambers~\cite{LHCb-DP-2012-002}.
The online event selection is performed by a trigger~\cite{LHCb-DP-2019-001},  which consists of a hardware stage (L0), based on information from the calorimeter and muon systems, followed by a software stage (HLT), which applies a full event reconstruction. The software stage employs a multivariate algorithm~\cite{BBDT,LHCb-PROC-2015-018} to identify secondary vertices consistent with the decay of a \bquark hadron.

Simulation is required to model the effects of the detector acceptance and the imposed selection requirements. In the simulation, $pp$ collisions are generated using \pythia~\cite{Sjostrand:2007gs} 
with a specific \lhcb configuration~\cite{LHCb-PROC-2010-056}.
Decays of unstable particles are described by \evtgen~\cite{Lange:2001uf}, in which final-state radiation is generated using \photos~\cite{davidson2015photos}. The interaction of the generated particles with the detector, and its response, are implemented using the \geant toolkit~\cite{Allison:2006ve, *Agostinelli:2002hh} as described in Ref.~\cite{LHCb-PROC-2011-006}. The underlying $pp$ interaction is reused multiple times, with an independently generated signal decay each time~\cite{LHCb-DP-2018-004}. 
Simulated $\Omegab$ and $\Xibm$ decays are generated for each of the data-taking years to account for different running conditions. These conditions are primarily related to changes in the hardware and software trigger thresholds, implemented to adapt to changes in the instantaneous luminosity.

\section{Data selection}
\label{sec:selection}

Signal $\Omegab$ and $\Xibm$ candidates are each reconstructed in pairs of kinematically similar decay modes. In the first pair of modes (hadronic modes), the decays $(\Omegab,\Xibm)\to(\Omegac\pim,\Xicz\pim)$ are used, where the $\Omegac$ and $\Xicz$ baryons are both reconstructed in the $p\Km\Km\pip$ final state. In the second set of modes ($\jpsi$ modes), the decays $(\Omegab,\Xibm)\to(\jpsi\Omegam,\jpsi\Xim)$ are used, with $(\Omegam,\Xim)\to(\Lz\Km,\Lz\pim)$, $\Lz\to p\pim$ and $\jpsi\to\mup\mun$. Hereafter, the notation $H_b$, $H_c$ and $H_s$ is used to refer to the $b$ baryons, the $c$ baryons and the $(\Xim,\Omegam)$ hyperons, respectively. 

A number of selections are imposed to reduce combinatorial background~\cite{Stripping}. All particles used to form the $H_b$ candidates are required to have trajectories that are significantly detached from all PVs in the event by requiring a large value of $\chisqip$. Here, $\chisqip$ is the difference between the $\chisq$ of the PV fit with and without the particle included. Its value is approximately equal to $({\rm IP}/\sigma_{\rm IP})^2$, and tends to be large for particles originating from $H_b$, $H_c$ or $H_s$ decay vertices, and small for particles originating from a PV. In addition, with the exception of the proton and pion from the $\Lz$ decay and the $\pim$ meson in the $\Xim$ decay, particle identification (PID) requirements are applied to the final-state particles to ensure compatibility with the identity assigned in the decay hypothesis. The reconstructed decay vertices for the $H_s$, $H_c$ and $H_b$ candidates are required to have good fit quality and have significant displacement from all PVs in the event.

For the hadronic modes, the $\Omegac$ ($\Xicz$) candidates are required to meet the mass requirements ${|M(p\Km\Km\pip)-m_{\Omegac(\Xicz)}|<20\,(15)\mevcc}$, corresponding to about 3 times the mass resolution. Here $M$ and $m$ refer to the reconstructed and known particle masses~\cite{PDG2026}, respectively. The $H_c$ decay times are required to satisfy $-0.20 < t_{\Xicz}< 1.24$ ps and $-0.20 < t_{\Omegac}< 2.08$ ps, where the lower limits are approximately twice the decay-time resolution, and the upper limits are about 8 times the known $H_c$ lifetimes~\cite{PDG2026}. To suppress background due to random combinations of hadrons, the $H_b$ candidate is required to have $\chisqip<16$.
Specific L0 and HLT triggers are required in order to have a well-defined trigger selection. Signal candidates in the hadronic modes are partitioned into two subsamples based on the information from the L0 hadron hardware trigger, ``triggered on signal'' (TOS) and ``triggered independently of the signal'' (TIS), as described in Ref.~\cite{LHCb-DP-2019-001}. Candidates are required to be in one of these two categories, which retains 96\% of the $H_b$ signal decays selected by any L0 trigger in the data. For the software trigger, the signal candidates are required to satisfy the requirements of the topological trigger~\cite{BBDT,LHCb-PROC-2015-018}. 

In the $\jpsi$ modes, the dimuon candidates are required to satisfy ${|M(\mup\mun)-m_{\jpsi}|<50\mevcc}$ and be significantly displaced from all PVs in the event. Due to the large lifetime of the $H_s$ and $\Lz$ baryons, both long (L) and downstream~(D) tracks are used in their reconstruction. Long tracks are reconstructed trajectories of charged particles with segments that start in the VELO detector and traverse the tracking system, while downstream tracks refer to charged particles that are reconstructed using only hits in the TT and T-stations (see Ref.~\cite{LHCb-DP-2008-001}, Fig. 10.1). Three combinations of long and downstream tracks are used to form the $H_s$ candidates, L(LL)$_{\Lz}$, L(DD)$_{\Lz}$ and D(DD)$_{\Lz}$, where the types of tracks forming the $\Lz$ are in parentheses. The $\Lz$ and $H_s$ candidates are required to satisfy the mass requirements, $|M(p\pim)-m_{\Lz}|<8\mevcc$ and 
$|M'(H_s)-m_{H_s}|<10\mevcc$. The usage of $M'(H_s)\equiv M(H_s)-M(p\pim)+m_{\Lz}$ improves the $H_s$ mass resolution by $\sim$30\% as compared to using $M(H_s)$.
A small contamination of misidentified $\Xim$ decays in the $\Omegam$ sample is removed by applying tighter PID requirements on the kaon from the $\Omegam$ decay when $|M'(\Lz\pim)-m_{\Xim}|<5\mevcc$, after the $\Km\to\pim$ mass replacement. The $H_b$ mass is determined in a kinematic fit~\cite{Hulsbergen:2005pu} where the $\jpsi$, $H_s$ and $\Lz$ masses are constrained to their known values~\cite{PDG2026}. 

The decay time $t$ of each $H_b$ candidate is computed with respect to the associated PV, taken to be the one in which $\chisqip$ is minimal. The $H_b$ decay time is required to be within the interval $0.4\leq t \leq 7.0$~ps. The lower decay-time requirement rejects events where the selection efficiency is decreasing rapidly as $t\to0$,  while the upper decay-time limit is sufficient to include about 98\% of the signal decays.

\subsection{Simulation corrections}
A set of corrections is applied to the simulation to account for imperfect modeling of the signal decays, the detector response and the relative luminosity for each year. The PID response of the final-state particles is calibrated using large samples of \mbox{$\Dstarp\to\Dz(\to\Km\pip)\pip$}, \mbox{${\Lb\to\Lc(\to\proton\Km\pip)\pim}$} and \mbox{$\jpsi\to\mup\mun$} decays as described in Ref.~\cite{LHCb-DP-2018-001}. Using the calibration data, the PID response for each particle type is parameterized as a function of its $p$, $\pt$ and the track multiplicity in the event~\cite{Poluektov_2015}. The PID response of each particle in simulated signal decays is updated to use the response obtained from these calibration data, chosen at random from the relevant PID probability distribution. The PID response for the kaon from the $\Omegam$ baryon uses the values from simulation, which is sufficient for the purposes of this measurement.

Both the hadronic and $\jpsi$ modes have weights applied to account for differences in the $(\pt,\eta)$ spectra of the $H_b$ baryon between simulation and data. The weights are obtained from the ratio of the background-subtracted $(\pt,\eta)$ spectrum of $\Xibm$ baryons in data\cite{Pivk:2004ty} to that in simulation. There are insufficient signal yields to perform the same weighting for the $\Omegab$ samples. Therefore, the $\Xibm$ $(\pt,\eta)$ weights are also used for the $\Omegab$ signal decays, and lead to good agreement between simulation and background-subtracted data. For the hadronic modes, the weights are determined separately for TOS and TIS events, and afterward, the simulations are weighted to have the same fractions of TOS and TIS events as in the background-subtracted data. The simulated samples are also weighted to account for differences between the $\Xibm$, $\Xicz$ and $\Omegac$ lifetimes in the simulation and the most recent measurements~\cite{LHCb-PAPER-2024-010,PDG2026}.
 
The $\Xicz$ and $\Omegac$ decays both show modest contributions of $\Kstarzb$ and $\LzStar$ resonance decays among the final-state particles, which are not included in the simulation. For the $\Xicz$ decay, weights are applied to the simulation based on the $M(\Km\pip)$ and $M(p\Km)$ spectra in semileptonic $\Xibm\to\Xicz\mun X$ decays~\cite{LHCb-PAPER-2019-008}, where $X$ represents any number of additional particles. For the $\Omegac$ decays, a weight for each $M(p\Km)$ and $M(\Km\pip)$ combination is applied, consisting of a non-relativistic Breit-Wigner line shape using the known masses and widths of the $\Kstarzb$ and $\LzStar$ resonances~\cite{PDG2026}. 

In addition to the weights described above, a correction is applied to the 2017 and 2018 simulations in the hadronic modes to account for small differences seen in the $\chisqip$ distributions between data and simulation. The weights are calibrated using the high signal yield control channel $\Lb\to\Lc\pim$, as discussed in Refs.~\cite{LHCb-PAPER-2024-010, LHCb-PAPER-2025-023}.

For the $\jpsi$ modes, it is found that the momentum asymmetries of the decay products of the $\Lz$ and $\Xim$ baryons, $\alpha\equiv(p_1-p_2)/(p_1+p_2)$, obtained from simulated decays do not fully reproduce the measured asymmetries in the data. Here, $p_1$ ($p_2$) is the total momentum of the larger (smaller) mass hadron in the decay, respectively. Weights are computed for the $\Lz\to p\pim$, $\Xim\to\Lz\pim$ and $\Omegam\to\Lz\Km$ parts of the signal decay from the ratio of the $\alpha$ distributions in background-subtracted data to that obtained from simulated signal decays. For the $\Lz\to p\pim$ portion of the decays, the data-simulation differences are consistent between the $\Xibm$ and $\Omegab$ decay modes, and the same set of weights $w_{\Lz}(\alpha)$ are applied to both simulation samples. For the $\Xim\to\Lz\pim$ and $\Omegam\to\Lz\Km$ decay asymmetries, only the former shows a statistically significant difference between background-subtracted data and simulation. Therefore, only weights $w_{\Xim}(\alpha)$ are applied to the $\Xibm\to\jpsi\Xim$ simulated decays.

\subsection{Multivariate selection}

To further suppress background from random combinations of reconstructed particles, a gradient-boosted decision tree algorithm~\cite{AdaBoost,Hocker:2007ht} (BDT) is employed. For the hadronic modes, the discriminating variables used in the BDT are identical to those used for the $\Xibm$ selection in Ref.~\cite{LHCb-PAPER-2024-010}. For the $\jpsi$ mode, the variables include the $H_b$, $H_s$ and $\Lz$ decay vertex-fit $\chisq$ values; $\cos(\theta_{\rm PV})$ for the $H_s$ and $\Lz$ candidates; the $H_s$ decay time; the $\pt$ of the $H_s$ decay products; the momenta of the $\pim$ from the $\Lz$ baryon and of the $\pim$ ($\Km$) meson from the $\Xim$ ($\Omegam$) baryon; the momentum asymmetries $\alpha$ of the $\Lz$ and $H_s$ baryons; the $\chisqip$ of the $\pim$ ($\Km$) from the $\Xim$ ($\Omegam$) baryon; and a PID variable for each muon~\cite{LHCb-DP-2018-001}. The variable $\theta_{\rm PV}$ is the angle between the momentum vector of the particle and the vector that connects the associated PV to its decay vertex.

For each mode, the BDT is trained using simulated signal decays with all corrections applied to represent the signal, and sidebands in the $H_b$ mass spectra to represent the background. The optimal BDT requirement for the $\Omegab$ candidates is chosen by maximizing the figure-of-merit $N_S/\sqrt{N_S+N_B}$, where $N_S\equiv\varepsilon N_0$ and $N_B$ are the estimated signal and background yields as a function of the BDT requirement. Here $N_0$ is the estimated signal yield with no BDT requirement, and $\varepsilon$ is the efficiency of a given selection on the BDT output, obtained from simulation. The background yield is obtained from the $H_b$ mass sidebands in data, scaled to reflect the expected background under the signal peak. The BDT requirements on the $\Xibm$ candidates are chosen to give approximately the same expected signal efficiency as for the $\Omegab$ decay mode. For the hadronic ($\jpsi$) modes, the efficiency of the BDT requirement is about 92\% (95\%), while suppressing the combinatorial background for the $\Omegab$ and $\Xibm$ candidates by factors of 25 and 6 (16 and 10), respectively.

After the BDT requirement, about 1\% of events have multiple candidates for both the $\jpsi$ and hadronic modes. To avoid potential double-counting of signal events, only the candidate with the largest BDT response is retained. 

\begin{figure}[tb]
\centering
\includegraphics[width=0.48\textwidth]{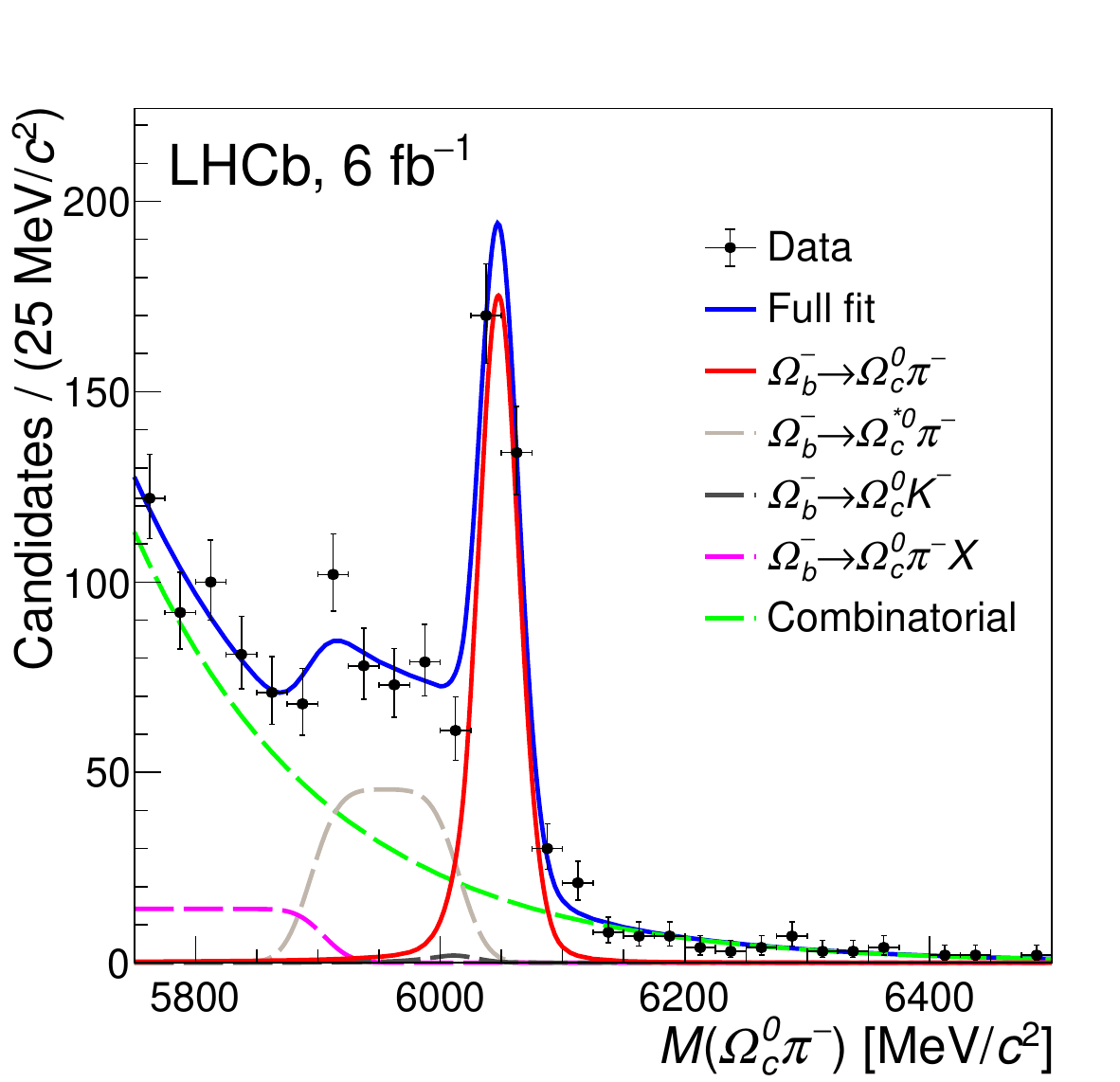}
\includegraphics[width=0.48\textwidth]{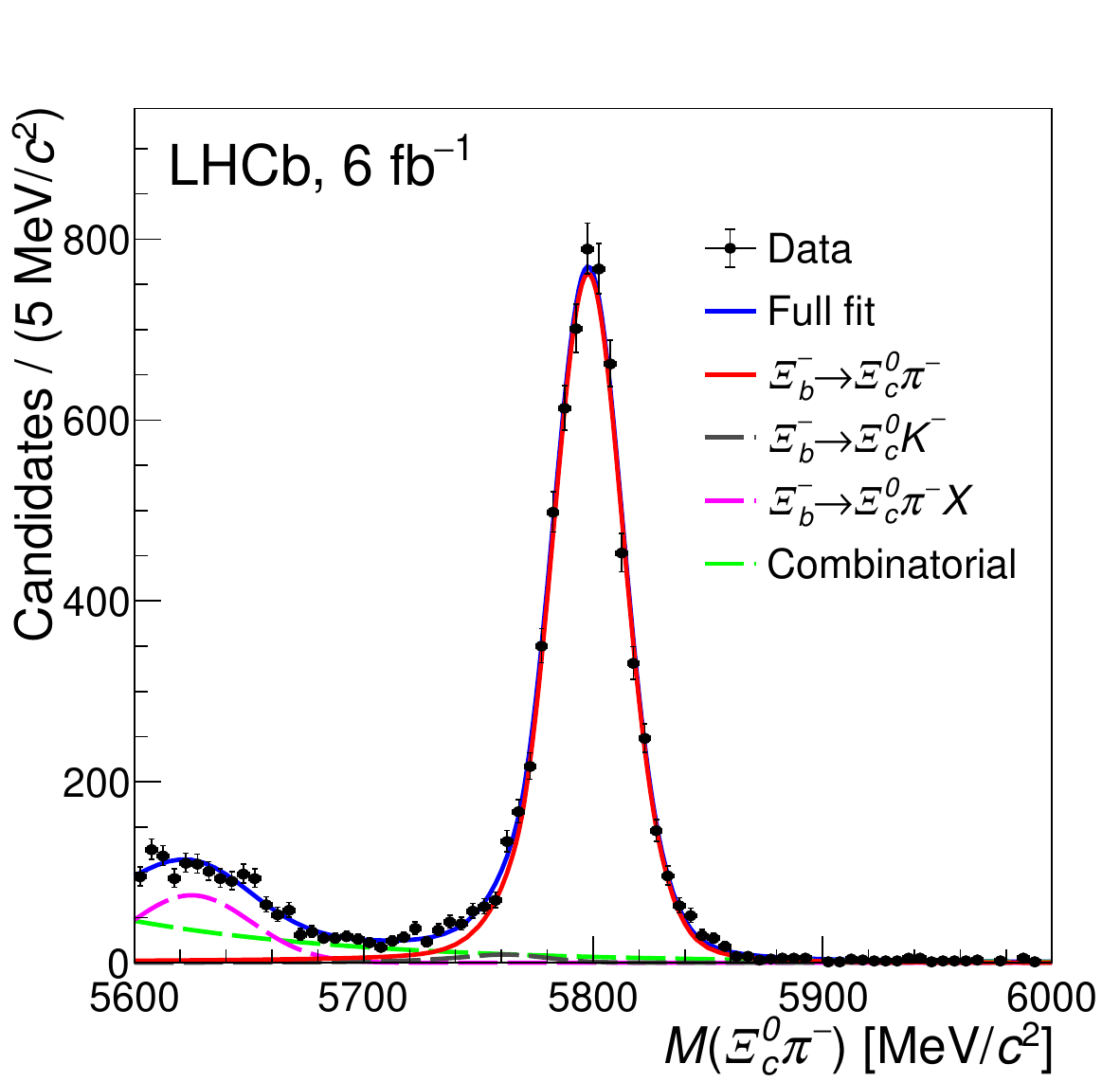}
\includegraphics[width=0.48\textwidth]{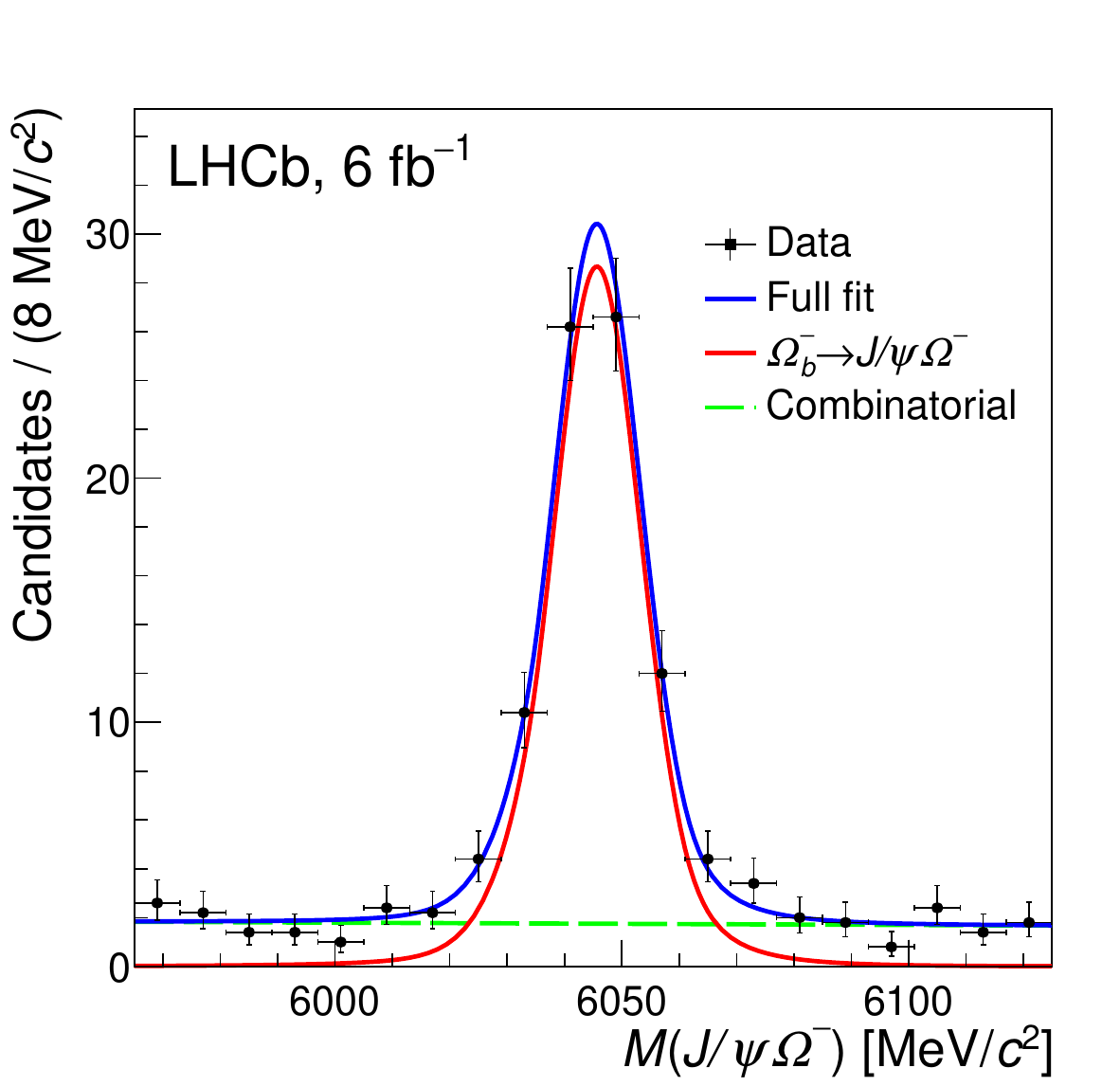}
\includegraphics[width=0.48\textwidth]{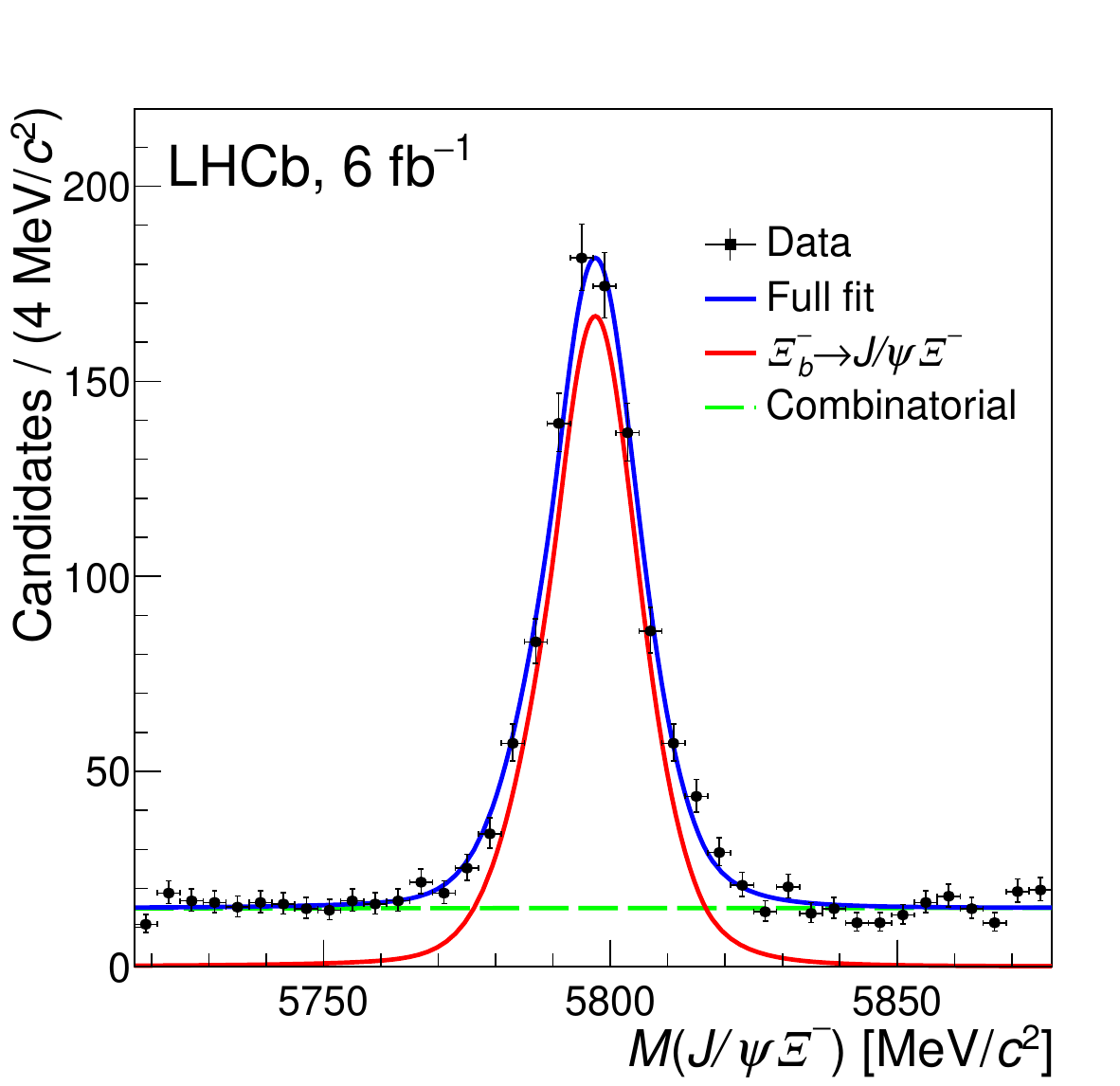}
\caption{\small{Mass spectra for (top left) \decay{\Omegab}{\Omegac\pim}, (top right) \decay{\Xibm}{\Xicz\pim}, (bottom left) \decay{\Omegab}{\jpsi\Omegam} and (bottom right) \decay{\Xibm}{\jpsi\Xim} candidates along with the results of the fits. Here, $\Omegares_c^{*0}$ refers to the $\Omegares_c(2770)^{0}$ baryon.}}
\label{fig:MassSpectraFinalSel}
\end{figure}
%%%%%%%%%%%%%%%%%%%%%%%%%%%%%%%%%%%%%%%%%
\subsection{Mass distributions}
%%%%%%%%%%%%%%%%%%%%%%%%%%%%%%%%%%%%%%%%%
The mass spectra for selected candidates are shown in Fig.~\ref{fig:MassSpectraFinalSel}, along with the results of unbinned extended maximum-likelihood fits. Each mass spectrum is described by the sum of a signal function and one or more background shapes. The signal-mass shapes are parameterized by the sum of two Crystal Ball functions~\cite{Skwarnicki:1986xj} with a common mean, and non-Gaussian tails on opposite sides of the signal peak. The signal shape parameters are fixed to the values obtained from simulation, except for the peak position and an overall mass resolution scale factor, which accounts for a small difference in the resolution between data and simulation. 

Several backgrounds pass the selection requirements of the hadronic modes. 
The background shapes associated with misidentified $H_b\to H_c\Km$ decays are obtained from simulation and fixed in the mass fits. The yield fraction $N(H_b\to H_c\Km)/N(H_b\to H_c\pim)$ is constrained with a Gaussian prior to $(1.5\pm0.3)\%$ based on known or estimated relative branching fractions~\cite{PDG2026} and efficiencies from simulation.  Partially reconstructed decays $H_b\to H_c\pim X$ with one or possibly more excluded particles appears at the low edge of the mass region under consideration. For the $\Xibm$ decay, a Gaussian shape with freely varying mean and width is used to describe this background. For the $\Omegab\to\Omegac\pim$ mode, this background is assumed to be predominantly $\Omegab\to\Omegac\rhom$ decays, and the shape is obtained from simulated
$\Omegab\to\Omegac\rhom$ decays~\cite{Cowan:2016tnm}. The shape is well described by a one-sided step function convolved with a Gaussian function, and is fixed in the fits to the data. The shape of the background from $\Omegab\to\Omegacstar\pim$ decays, with $\Omegacstar\to\Omegac\gamma$, is obtained from simulation~\cite{Cowan:2016tnm}. The $\Omegac\pim$ mass spectrum is parameterized by a two-sided step function convolved with a Gaussian function that accounts for the mass resolution in data, as obtained from the $\Omegab\to\Omegac\pim$ signal peak. The combinatorial background shapes in  the signal and normalization modes are each described by an exponential function with a freely varying shape parameter.

In the $\jpsi$ modes, there are no known partially reconstructed or misidentified decays that would produce any significant structure in the candidate mass spectra within the fit regions. Thus, an exponential function is used to describe the background shapes in the $\Omegab$ and $\Xibm$ mass spectra with freely varying shape parameters.

The fitted signal yields in the hadronic modes are $318\pm21$ and $6552\pm87$ for the $\Omegab$ and $\Xibm$ decay modes, respectively, which are 5.1 and 4.7 times larger than the Run 1 analysis~\cite{LHCb-PAPER-2016-008}. For the $\jpsi$ modes, the $\Omegab$ and $\Xibm$ signal yields are $373\pm22$ and $2234\pm57$, respectively, corresponding to relative increases of 6.4 and 7.1 with respect to the Run 1 analysis~\cite{LHCb-PAPER-2014-010}. The larger increases in signal yields for the $\jpsi$ modes are due to less stringent selection requirements in this analysis compared to those of Run 1. The mass resolution scale factors are $1.12\pm0.02$ and $1.10\pm0.03$ for the hadronic and $\jpsi$ modes, respectively, which are consistent with other analyses of $b$-hadron decays in LHCb.

%%%%%%%%%%%%%%%%%%%%%%%%%%%%%%%%%%%%%%%%%%
\section{Determination of $r_{\tau}$}
%%%%%%%%%%%%%%%%%%%%%%%%%%%%%%%%%%%%%%%%%%
The determination of $r_{\tau}$ requires the ratio of signal yields and the ratio of efficiencies, as shown in Eq.~\ref{eq:RT}. The yields $N[\Omegab\to f_s](t)$ and $N[\Xibm\to f_n](t)$ are determined in specific decay-time bins. The bin boundaries are chosen with input from pseudoexperiments such that any bias associated with the fit for $r_{\tau}$ (due to small sample sizes) must be small compared to other systematic uncertainties. For the hadronic modes, the time bins are $[0.4,\,1.0]\ps$, $[1.0,\,1.7]\ps$, $[1.7,\,2.5]\ps$, $[2.5,\,3.4]\ps$, $[3.4,\,4.8]\ps$, and $[4.8,\,7.0]\ps$. For the $\jpsi$ modes the ranges are $[0.4,\,0.7]\ps$, $[0.7,\,1.0]\ps$, $[1.0,\,1.9]\ps$, $[1.9,\,3.0]\ps$, $[3.0,\,5.1]\ps$, and $[5.1,\,7.0]\ps$. The different boundaries are a consequence of the different decay-time acceptances of the hadronic and $\jpsi$ modes. 

The signal and background mass shapes in each decay-time bin are the same as described for the full sample.
The $H_b$ signal yields in each decay-time bin are obtained through a simultaneous extended unbinned maximum-likelihood fit across all decay time bins. The $H_b$ fits are performed independently, since they do not share any fit parameters. For each $H_b$ mode, some of the parameters are shared across decay-time bins, including the yield ratios $N(\Omegab\to\OmegacStar\pim)/N(\Omegab\to\Omegac\pim)$,  $N(\Omegab\to\Omegac\pim X)/N(\Omegab\to\Omegac\pim)$, and $N(\Omegab\to\Omegac\Km)/N(\Omegab\to\Omegac\pim)$, $N(\Xibm\to\Xicz\Km)/N(\Xibm\to\Xicz\pim)$, the peak position of the $H_b$ signal component, and the mass resolution scale factor. Taking these parameters as common to all decay-time bins is validated using a large sample of $\Lb\to\Lc\pim X$ decays. Due to the nearly identical kinematics, the resolution scale factors for the $\Omegab$ decays are constrained with a Gaussian prior to the values obtained in the $\Xibm$ mass fits. The width is set to be three times the uncertainty found in the $\Xibm$ mass fit, in order to allow for some reasonable variation. The remaining parameters, such as the exponential background shape parameters and the combinatorial background and signal yields, are freely varied in the fit in each decay-time bin.

\begin{figure}[tb]
\centering
\includegraphics[width=0.48\textwidth]{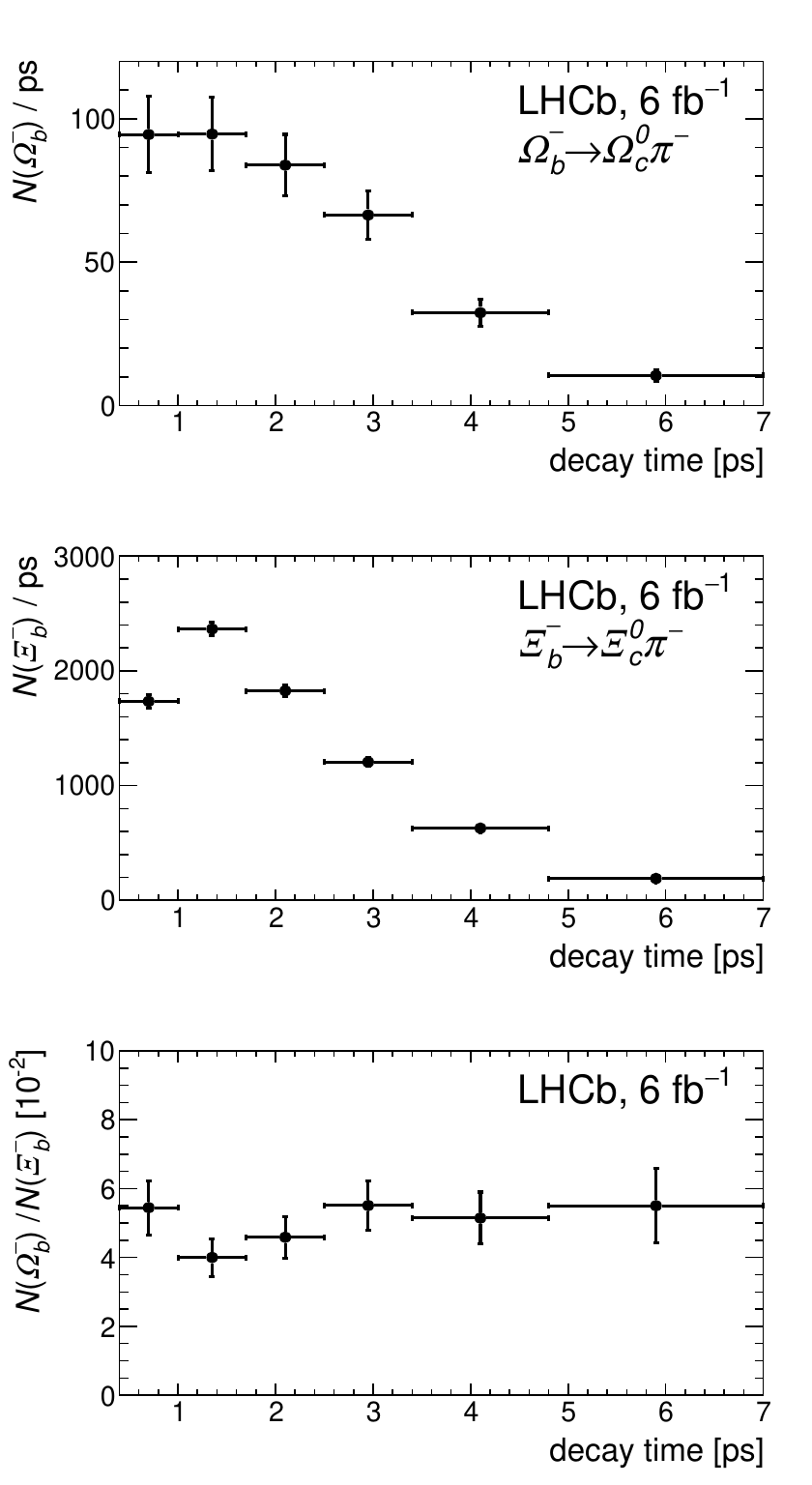}
\includegraphics[width=0.48\textwidth]{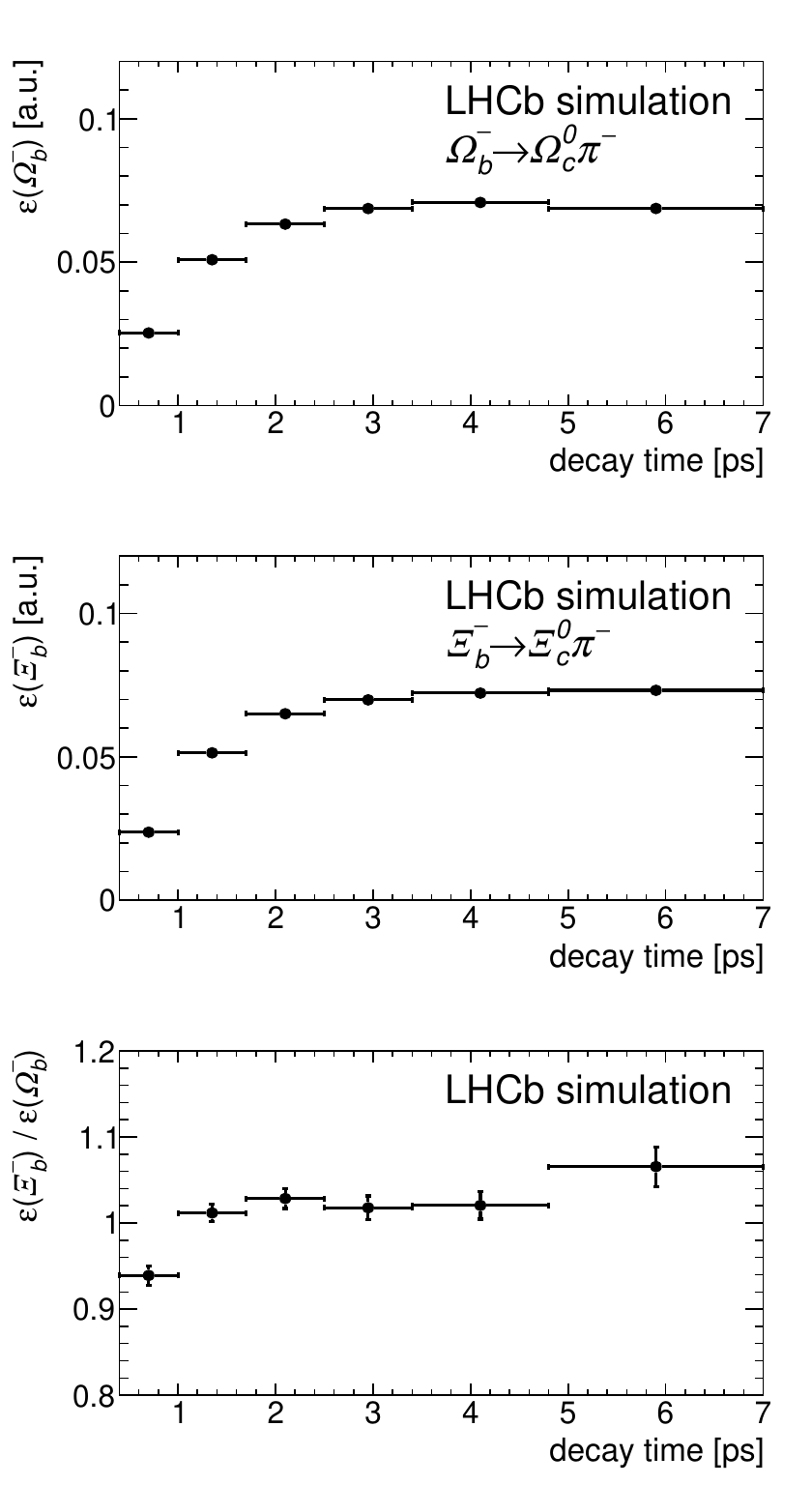}
\caption{\small{(Left) Signal yields in the (top) \decay{\Omegab}{\Omegac\pim} mode, and (middle) \decay{\Xib}{\Xic\pim} mode, along with (bottom) the ratio of signal yields ($\Omegab/\Xibm$). (Right) Efficiencies for the (top) \decay{\Omegab}{\Omegac\pim} mode, and (middle) the \decay{\Xib}{\Xic\pim} mode, and (bottom) the ratio of efficiencies ($\Xibm/\Omegab$), after all simulation weights are applied.}}
\label{fig:YieldEffSumHad}
\end{figure}
\begin{figure}[tb]
\centering
\includegraphics[width=0.48\textwidth]{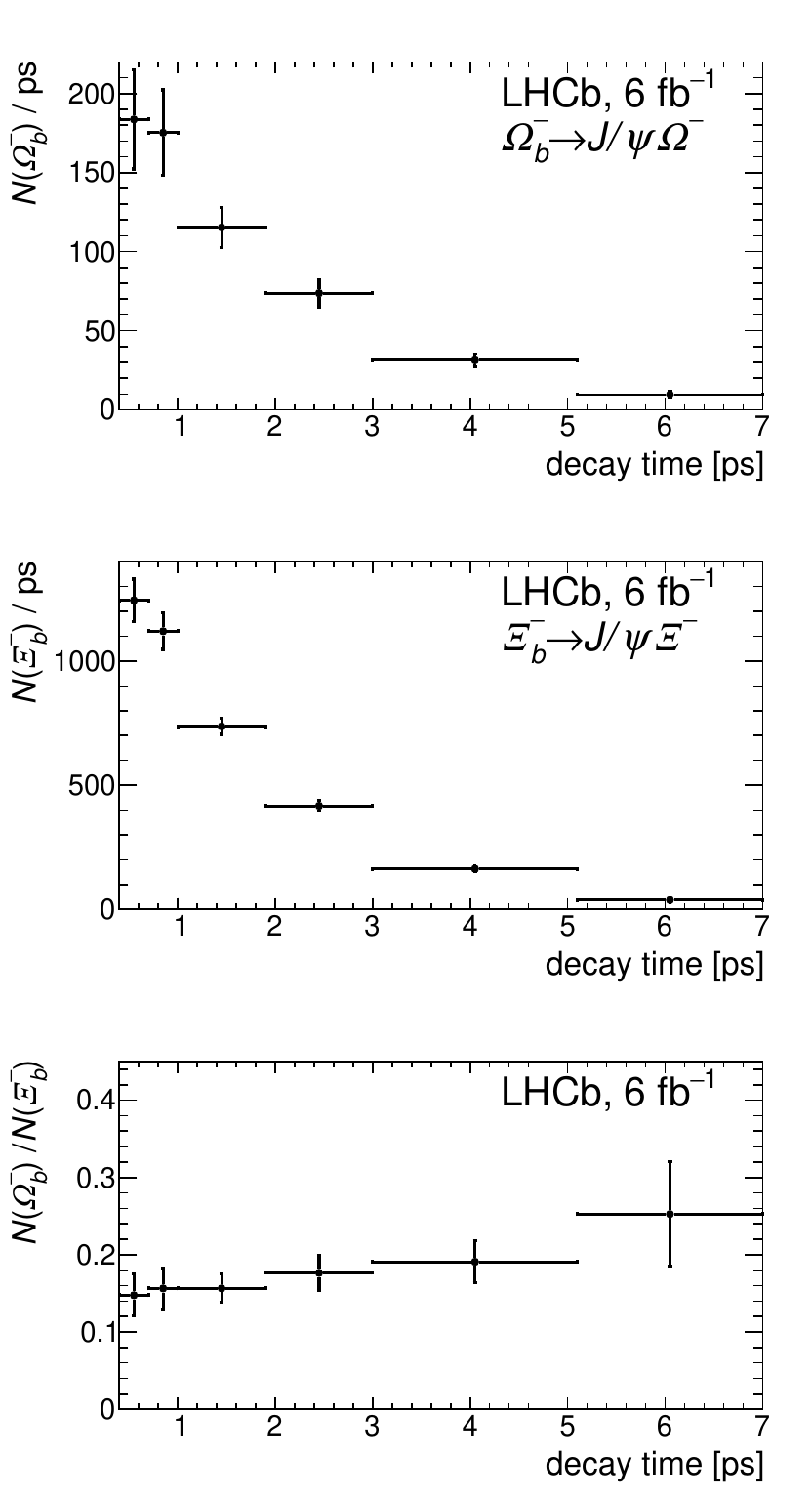}
\includegraphics[width=0.48\textwidth]{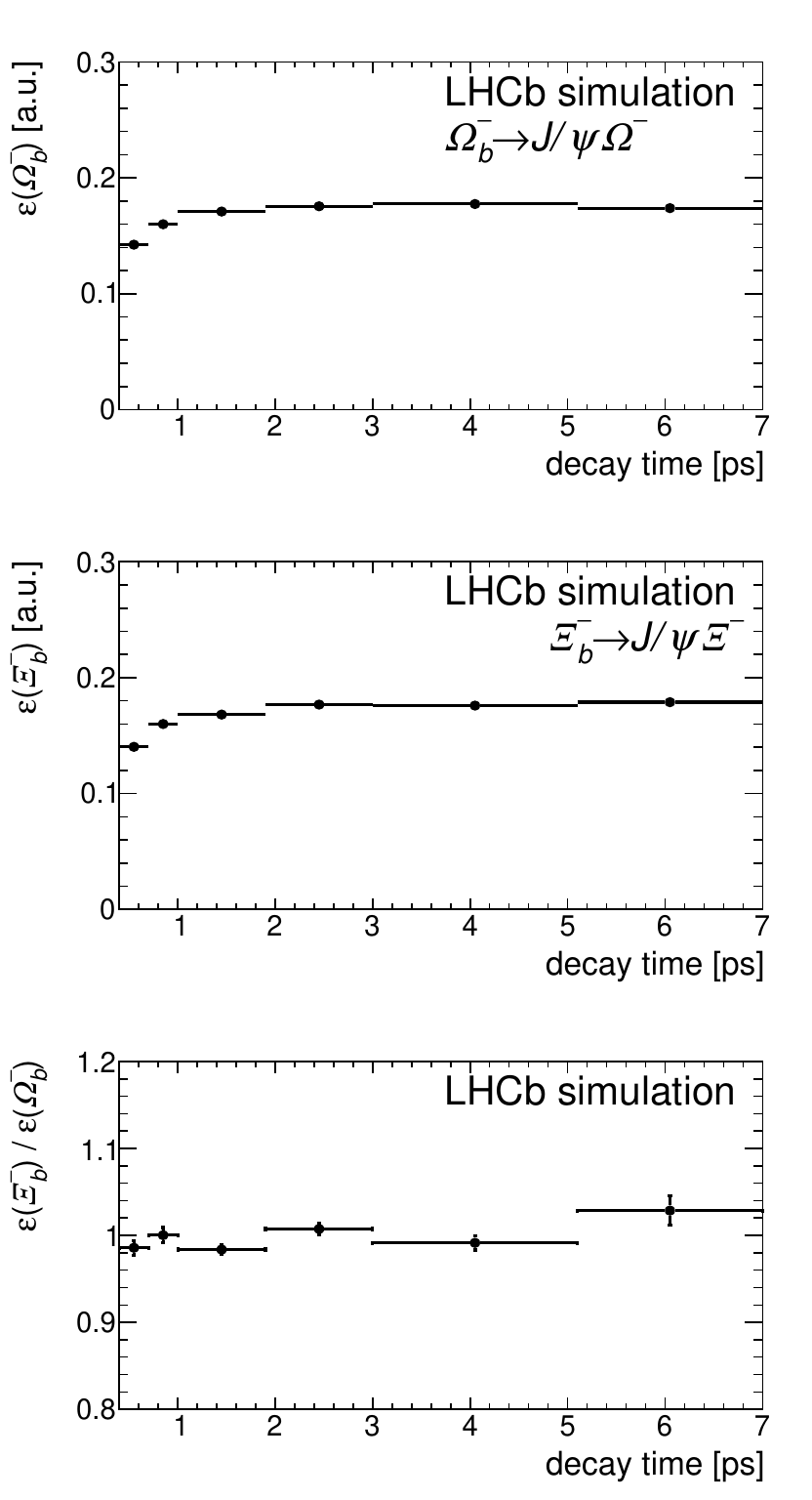}
\caption{\small{(Left) Signal yields in the (top) \decay{\Omegab}{\jpsi\Omegam} mode, and (middle) \decay{\Xibm}{\jpsi\Xim} mode, and (bottom) the ratio of signal yields ($\Omegab/\Xibm$). (Right) Efficiencies for the (top) \decay{\Omegab}{\jpsi\Omegam} mode, and (middle) the \decay{\Xibm}{\jpsi\Xim}  mode, and (bottom) the ratio of efficiencies ($\Xibm/\Omegab$), after all simulation weights are applied.}}
\label{fig:YieldEffSumJPsi}
\end{figure}
The yields and efficiencies of $\Omegab\to\Omegac\pim$ and $\Xibm\to\Xicz\pim$ decays as a function of the decay time are shown in Fig.~\ref{fig:YieldEffSumHad}, along with the ratio of yields and efficiencies. The efficiencies are computed with all weights applied to the simulation. The relative efficiency and yield are close to flat for $t>1.0$ ps. For $t<1.0$~ps, the $H_c$ decay time becomes an important contribution to the value of $\chisqip$, and the larger $\Omegac$ baryon lifetime compared to that of the $\Xicz$ baryon leads to an increase in the yield ratio $N(\Omegab)/N(\Xibm)$ and a decrease in the efficiency ratio $\eff(\Xibm)/\eff(\Omegab)$. The analogous distributions for the $\jpsi$ modes are shown in Fig.~\ref{fig:YieldEffSumJPsi}. 
The IP-dependent selections in the $\jpsi$ modes are less stringent than those of the hadronic modes due to the much lower rate of promptly produced muons than hadrons in $pp$ collisions. This results in a larger efficiency to reconstruct $H_b$ baryons with small decay time in the $\jpsi$ modes as compared to the hadronic modes.
Due to the very similar topologies of the $\Omegab\to\jpsi\Omegam$ and $\Xibm\to\jpsi\Xim$ decays and the nearly identical selection requirements, the ratio of efficiencies is consistent with being flat with decay time.

\begin{figure}[tb]
\centering
\includegraphics[width=0.48\textwidth]{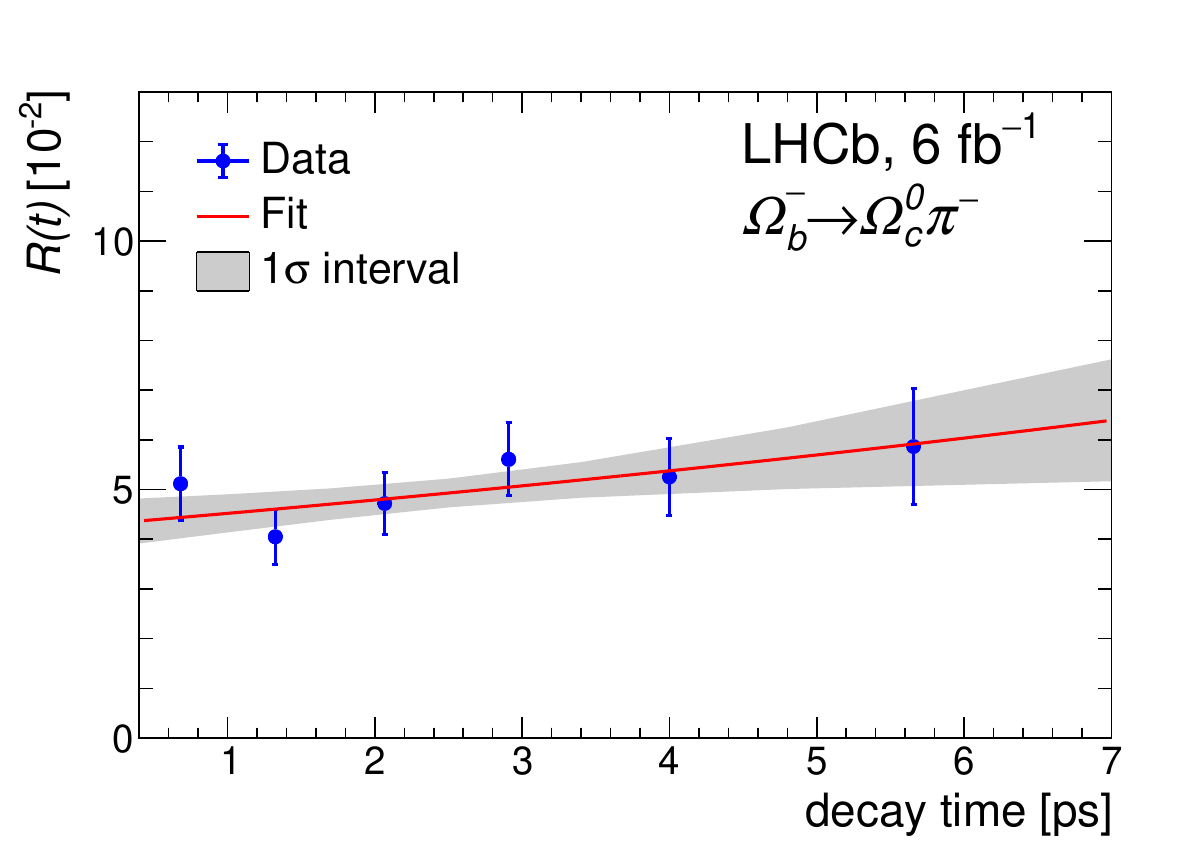}
\includegraphics[width=0.48\textwidth]{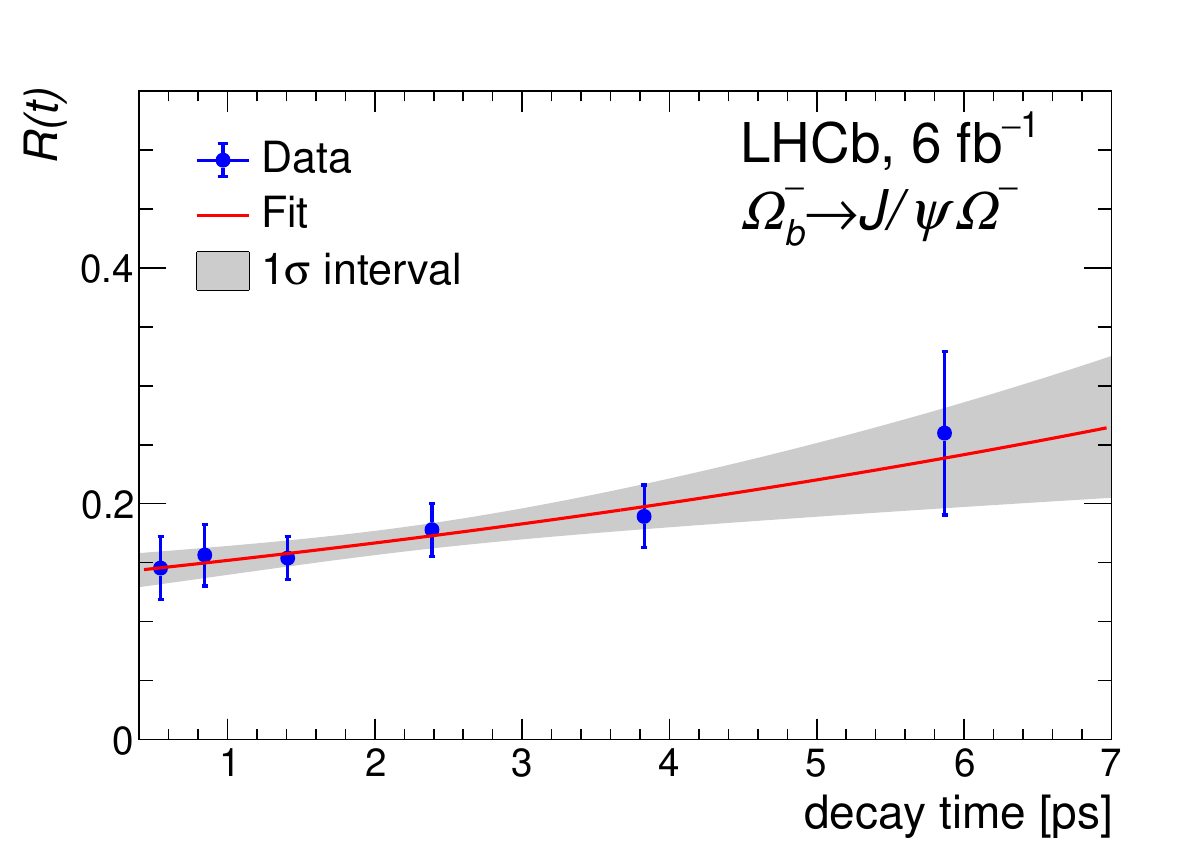}
\includegraphics[width=0.48\textwidth]{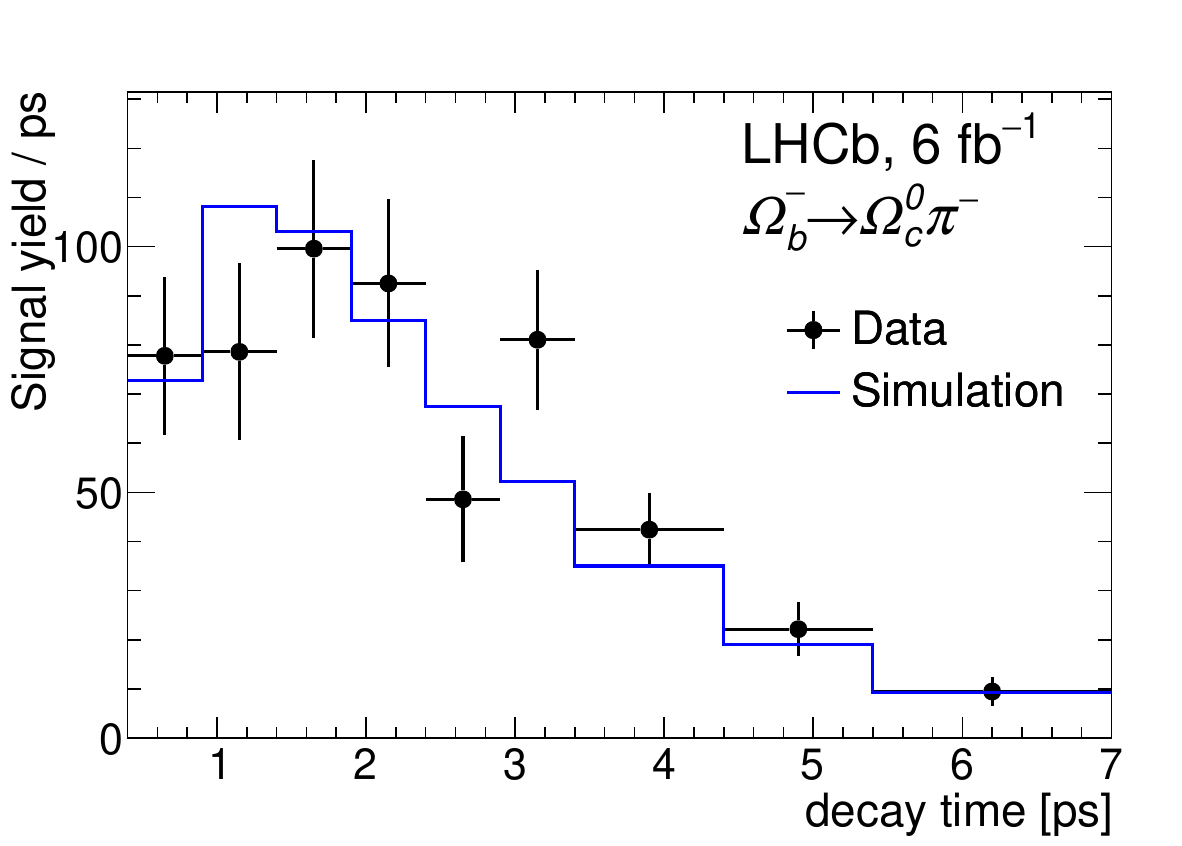}
\includegraphics[width=0.48\textwidth]{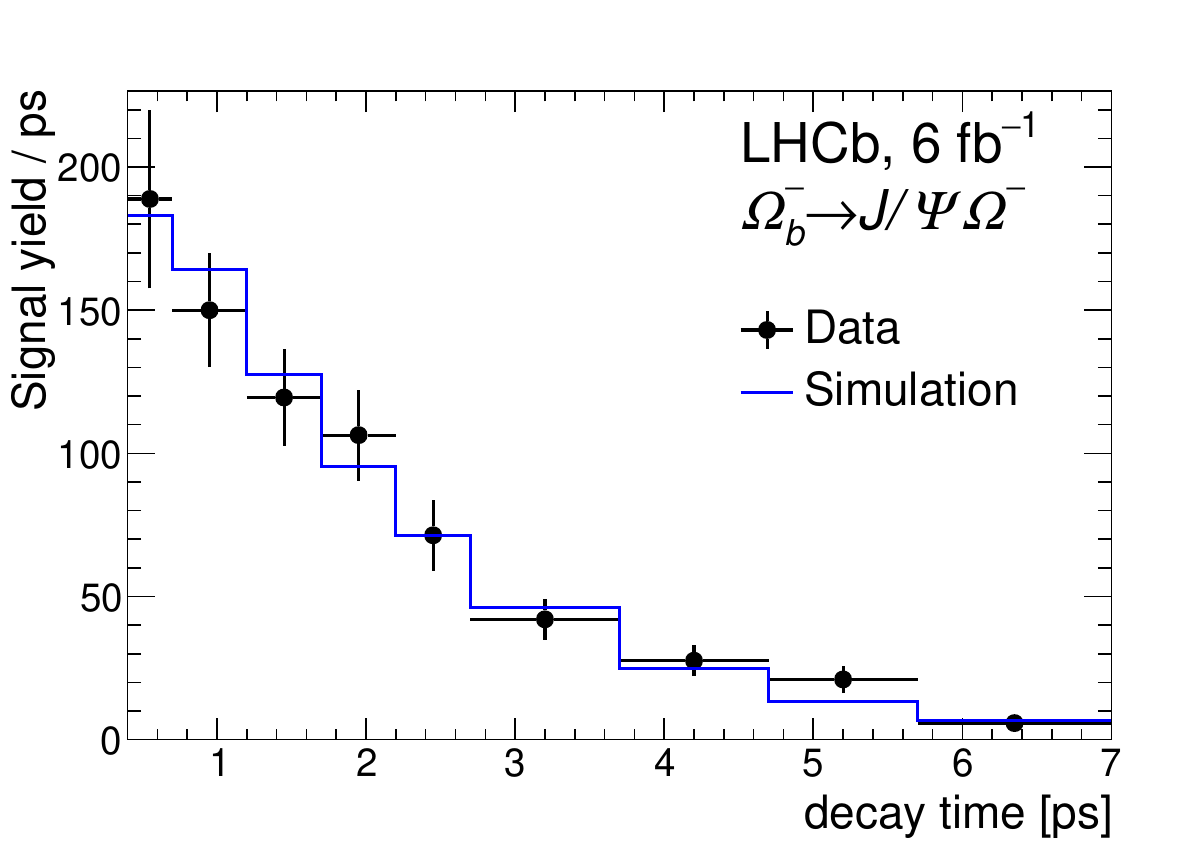}
\caption{\small{Corrected yield ratio $R(t)$ as a function of the decay time for the (top left) hadronic modes and (top right) $\jpsi$ modes. The red lines show the fit, and the gray band shows the 68\% confidence level intervals. The bottom distributions show the corresponding background-subtracted $\Omegab$ decay-time spectra~\cite{Pivk:2004ty} in data compared to the simulation, weighted to represent the best fit ${\Omegab}$ lifetimes of $\tau_{\Omegab}^{\rm had}=1.74$~ps and $\tau_{\Omegab}^{\jpsi}=1.85$~ps. 
}}
\label{fig:CorrYieldRatio}
\end{figure}

The product of the yield and efficiency ratios shown in the bottom of Figs.~\ref{fig:YieldEffSumHad} and ~\ref{fig:YieldEffSumJPsi} gives the efficiency-corrected yield ratios $R(t)$ for each mode, as shown in Fig.~\ref{fig:CorrYieldRatio}\,(top). The points are placed along the time axis at the average value within the bin as described in Ref.~\cite{Lafferty}. Fits to $R(t)$ for the hadronic and $\jpsi$ modes with the exponential function shown in Eq.~\ref{eq:RT} result in the values \mbox{$\lambda^{\rm had}=(5.8\pm4.1)\times10^{-2}$~ps$^{-1}$} and \mbox{$\lambda^{\jpsi}=(9.3\pm4.5)\times10^{-2}$ ps$^{-1}$}, where the uncertainties are statistical only.. The relative lifetimes are readily computed from Eq.~\ref{eq:rtau}, resulting in the corresponding values \mbox{$r_{\tau}^{\rm had}=1.100\pm0.079$} and \mbox{$r_{\tau}^{\jpsi}=1.172\pm0.102$}, where the uncertainties are statistical only. The measured $r_{\tau}$ values are consistent with each other. Using the recently measured $\Xibm$ lifetime of $1.579\pm0.023$~ps~\cite{LHCb-PAPER-2024-010,PDG2026}, the $\Omegab$ lifetimes are measured to be \mbox{$\tau_{\Omegab}^{\rm had}=1.74\pm0.12$~ps} and \mbox{$\tau_{\Omegab}^{\jpsi}=1.85\pm0.16$\ps}.

Figure~\ref{fig:CorrYieldRatio}\,(bottom) shows an overlay of the decay-time spectrum of $\Omegab$ signal decays obtained using the \sPlot method~\cite{Pivk:2004ty}, and that of simulated decays weighted to represent the best fit $\Omegab$ lifetimes given above. It is seen that the decay time spectrum in simulation using the measured lifetime describes well the decay-time spectrum in data.

\section{Systematic uncertainties}

A number of sources of systematic uncertainty are listed in Table~\ref{tab:Syst} along with their associated values. For each of the sources, unless otherwise indicated, the systematic uncertainty is assessed by making a change to the default method, and assigning the relative change in $r_{\tau}$ with respect to the baseline value as the systematic uncertainty. The total systematic uncertainty is obtained from the quadrature sum of the values from each source.

\begin{table*}[t]
\begin{center}
\caption{\small{Systematic and statistical uncertainties on $r_{\tau}$ for the hadronic and $\jpsi$ modes. The total systematic uncertainty is the quadrature sum of the values from each source.}}
\begin{tabular}{lcc}
\hline\hline
                        &   \multicolumn{2}{c}{Value (\%)} \\
Source                  & Hadronic  &  $\jpsi$ \\
\hline
Simulated sample size     &   0.65   &  0.46 \\
Signal shapes             &   0.15   &  0.11 \\    
Background shapes         &   0.47   &  0.69 \\  
Truth matching            &   0.36   &  0.04 \\  
Muon $\chisqip$           &     --    &  0.07 \\
$r_{\tau}$ fit bias       &   0.20   &  0.38 \\   
BDT requirement           &   0.03   &  0.05 \\  
$\Xibm$ lifetime          &   0.16   &  0.25 \\
\hline
Total systematic          &   0.93   &  0.96\\
\hline
Statistical               &   7.14   &  8.75 \\
\hline\hline
\end{tabular}
\label{tab:Syst}
\end{center}
\end{table*}

The uncertainty due to finite simulation sample sizes is obtained by performing 5000 alternative fits for $r_{\tau}$, where in each fit the relative efficiency in each bin is fluctuated by its uncertainty according to a Gaussian prior. The standard deviation of the 5000 alternative fits is assigned as the systematic uncertainty.

The uncertainty due to the assumed signal shape is quantified by using the sum of a Bukin~\cite{Verkerke:2003ir} and Gaussian function with a common peak value for the signal function to fit the data. For the hadronic modes, a second Gaussian function is required to account for the low mass tail due to final-state radiation from the charged particles. 

Sensitivity to the choice of the combinatorial background function is investigated by using other smooth shapes, and is found to be negligible in both the $\jpsi$ and hadronic modes. For the $\Omegab\to\jpsi\Omegam$ mode, in a wider mass region from 5885--6125\mevcc, there is potentially a small step in the background level at $\sim$5980\mevcc. Although there are no known backgrounds that could produce this small step, it is assumed to be some unknown background and its effect on $r_{\tau}$ is studied by adding a step function convolved with a Gaussian function to model its shape. 

For the hadronic modes, mismodeling of the partially reconstructed backgrounds in both the $\Xibm$ and $\Omegab$ baryon decays are considered. The baseline $\Xibm$ mass fits do not include a $\Xibm\to\Xicprime\pim$ with $\Xicprime\to\Xicz\gamma$ component. In an alternative fit, a $\Xicprime$ component is added with a shape obtained using simulated $\Xibm\to\Xicprime\pim$ decays~\cite{Cowan:2016tnm}, leading to a change in $r_{\tau}$ of 0.30\%. For the ${\Omegab\to\OmegacStar\pim}$ component, two modifications are made with respect to the baseline fit. First, the resolution scale factor applied to the Gaussian function of the ${\Omegab\to\OmegacStar\pim}$ shape is increased from 1.12 to 1.24. Secondly, usage of a common fraction ${\bar{f}=N(\Omegab\to\OmegacStar\pim)/N(\Omegab\to\Omegac\pim)}$ for all decay time bins is replaced by the six fractions $f_i=\bar{f}+\Delta f_i$. The parameters $\bar{f}$ and $\Delta f_3$ are freely varied, while the other $\Delta f_i$ are constrained with a Gaussian prior with a mean of 0 and a width of 3\% of $\bar{f}$, based on the measured variation of the fraction $N(\Lb\to\Lc\pim X)/N(\Lc\to\Lc\pim)$ with decay time. This alternative fit results in a change in $r_{\tau}$ of 0.36\%. The $\Xibm$ and $\Omegab$ background uncertainties are added in quadrature.

For the efficiency determination, each reconstructed signal decay in the simulation is required to be truth-matched to the generated signal decay. In a few percent of cases, this association can fail if the fraction of matched hits between the reconstructed track and the true particle does not exceed a minimum threshold. The potential impact is assessed by including these unmatched decays in the efficiency determination.

For the $\jpsi$ modes, the $\chisqip$ corrections are not applied. To investigate any potential effects due to data-simulation differences in $\chisqip$ for the muons, the baseline requirement is increased from $\chisqip>3$ to $\chisqip>4.5$ only in the simulation, and the efficiencies are re-evaluated. The change of 0.07\% is assigned as a systematic uncertainty.

The time bins are chosen with the aim of balancing the signal yields in each decay-time bin in order to maintain a small fit bias. The fit bias is estimated using 1000 $r_{\tau}$ fits, where in each fit the number of $\Omegab$ and $\Xibm$ signal decays is the same as that observed in the data, with decay times selected randomly according to the decay-time spectrum in the weighted simulation. The resulting fit bias is taken to be the deviation in the mean $r_{\tau}$ from the baseline value.

The background-subtracted BDT spectrum in data is well modeled by the weighted simulation; however, the agreement is not perfect. To estimate a potential impact of the BDT requirement on the measurement of $r_{\tau}$, the BDT spectra in the simulation are weighted to match the background-subtracted BDT spectra in data. The relative difference between the BDT-weighted and the baseline value of $r_{\tau}$ is assigned as a systematic uncertainty.

The systematic uncertainty in $r_{\tau}$ due to the limited precision of the $\Xibm$ baryon lifetime is obtained by propagating the uncertainty in Eq.~\ref{eq:rtau}. The total relative systematic uncertainty on $r_{\tau}$ is 0.93\% and 0.96\% for the hadronic and $\jpsi$ modes, respectively; these values are substantially lower than the corresponding statistical uncertainties of 7.14\% and 8.75\%.

\section{Results and summary}
\label{sec:Conclusion}
In summary, a $pp$ collision data sample corresponding to an integrated luminosity of 6\invfb and collected by the LHCb experiment is used to measure the ratio of lifetimes of the $\Omegab$ baryon relative to that of the $\Xibm$ baryon. The measured lifetime ratios are
\begin{align*}
    r_{\tau}^{\rm had}&=1.100\pm0.079\pm0.010, \\
    r_{\tau}^{\jpsi}&=1.172\pm0.102\pm0.011,
\end{align*}    
\noindent using the hadronic ($\Omegab\to\Omegac\pim$ and $\Xibm\to\Xicz\pim$) and the $\jpsi$ ($\Omegab\to\jpsi\Omegam$ and $\Xibm\to\jpsi\Xim$) decay modes. The uncertainties are statistical and systematic.

These results are consistent with each other, and about a factor of two more precise than earlier measurements from LHCb~\cite{LHCb-PAPER-2014-010, LHCb-PAPER-2016-008} using 3\invfb of $pp$ collision data collected at $\sqrt{s}=7$ and 8\tev. Those previous measurements are combined with those presented here. Among the systematic uncertainties, only the simulated sample sizes and muon $\chisqip$ uncertainties are taken to be uncorrelated, and the rest are taken to be 100\% correlated for sources in common. The average values for Run 1 and Run 2 are 
\begin{align*}
    r_{\tau}&=1.109\pm0.055\pm0.010 , \\
    \tau_{\Omegab}&=1.751\pm0.089\pm0.022~{\rm ps},
\end{align*}
where the value $\tau_{\Xibm}=1.579\pm0.018\pm0.014\ps$~\cite{LHCb-PAPER-2024-010,PDG2026} has been used. In computing \mbox{$\tau_{\Omegab}=r_{\tau}\cdot\tau_{\Xibm}$}, the $\Xibm$ samples used in these measurements have a large overlap, which leads to correlated uncertainties between $r_{\tau}$ and $\tau_{\Xibm}$. The correlation coefficient between the statistical uncertainties is found to be $-0.025$, while for the systematic uncertainties it is $-0.148$. These correlations are included in the final value of $\tau_{\Omegab}$. The results presented here supersede the earlier Run 1 measurements.

\begin{figure}[tb]
\centering
\includegraphics[width=0.98\textwidth]{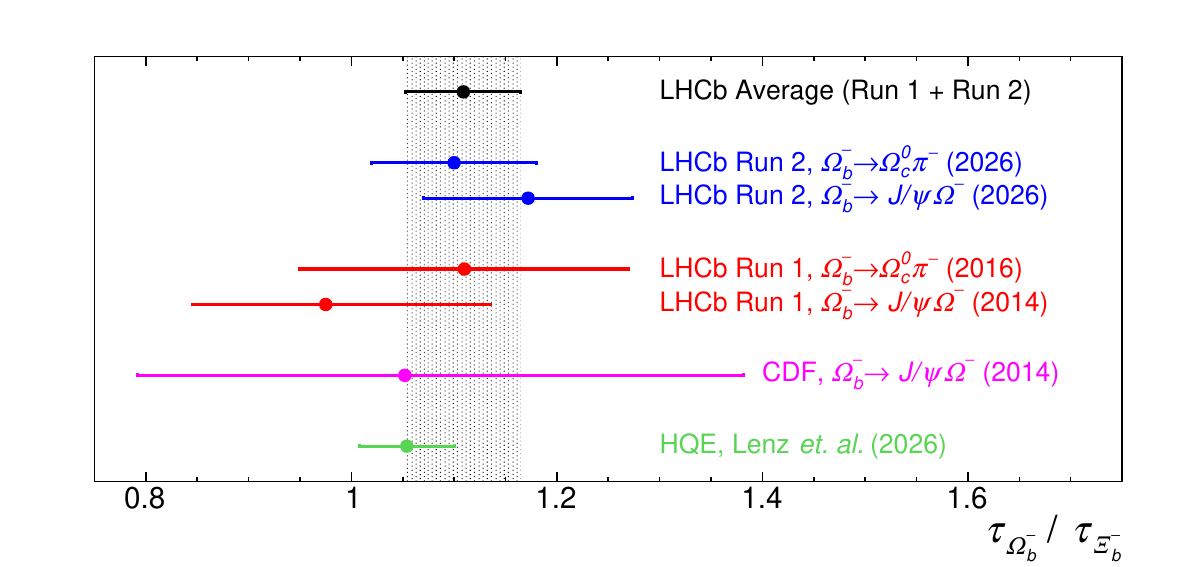}
\caption{\small{Comparison of the Run 1 and Run 2 measured lifetime ratios, $\tau_{\Omegab}/\tau_{\Xibm}$, along with that of the CDF measurement (using the \decay{\Omegab}{\jpsi\Omegam} decay). The average of all LHCb measurements is shown at the top, and the HQE prediction~\cite{lenz2026predictionsbbaryonlifetimesnnloqcd} at the bottom.  
}}
\label{fig:LifetimePlot}
\end{figure}

Figure~\ref{fig:LifetimePlot} compares the $r_{\tau}$ values for 
the two Run 1 and Run 2 LHCb measurements and the average value. The CDF measurement using $\Omegab\to\jpsi\Omegam$ is also shown. For the CDF and LHCb Run 1 $\Omegab\to\jpsi\Omegam$ measurements, the absolute lifetimes are cast into lifetime ratios using the 
$\Xibm$ lifetime by LHCb~\cite{LHCb-PAPER-2024-010}. All of the measurements are compatible with the LHCb average. The HQE prediction for the ratio is ${r_{\tau}^{\rm HQE}=1.054\pm0.046}$~\cite{lenz2026predictionsbbaryonlifetimesnnloqcd} and is also shown in the figure. The measured and predicted values are consistent within one standard deviation. 

The $s\to u\Wm$ transitions within the $H_b$ baryons contribute to the total decay widths of the $\Omegab$ and $\Xibm$ baryons, and therefore are included in the measured value of $r_{\tau}$. However, such contributions are not treated within the HQE framework. The $\Xibm\to\Lb\pim$ decay has been observed with \mbox{${\cal{B}}(\Xibm\to\Lb\pim)=(0.89\pm0.10\pm0.07\pm0.29)\%$}~\cite{LHCb-PAPER-2023-015}, which is consistent with a range of predictions~\cite{PhysRevD.105.094011,Cheng:2016jhep,NIU2022136916,PhysRevD.93.034020,FALLER2015653,PhysRevD.106.093005}. In contrast, the contribution of the decay $\Omegab\to\Xibz\pim$ to $\Gamma_{\!\Omegab}$ is predicted to be much smaller~\cite{FALLER2015653,PhysRevD.105.094011,PhysRevD.106.093005}, below 0.1\%. In both $\Xibm$ and $\Omegab$ baryon decays, the semileptonic transitions $s\to u(\Wm\to\ellm\neulb)$ are expected to be much smaller than those of the pionic modes $s\to u(\Wm\to\pim)$~\cite{FALLER2015653,OmegabSL}. As a result, if the $s\to u\Wm$ contributions are considered as a correction to $r_{\tau}^{\rm HQE}$, it would shift the value upward by about 1\%. Compared to the current level of precision on $r_{\tau}$, the effect is rather small. Both the theoretical and experimental measurements will benefit from further reductions in the uncertainties in the future, which will allow for a more stringent test of the HQE prediction.

% Do not include this in any draft (just for information in the template)
%\input{LHCbInternal/acknowledgements_template}
% Comment this in for paper drafts; do not include this in analysis note, conference and figure reports
\section*{Acknowledgements}
%
% These Acknowledgements valid from 16/06/2026
%
\noindent 
We thank Alexander Lenz, Maria Laura Piscopo and Aleksei Rusov for fruitful exchanges on the HQE framework, and for producing updated HQE predictions in advance of this publication. 
We express our gratitude to our colleagues in the CERN
accelerator departments for the excellent performance of the LHC. We
thank the technical and administrative staff at the LHCb
institutes.
We acknowledge support from CERN and from the national agencies:
ARC (Australia);
CAPES, CNPq, FAPERJ and FINEP (Brazil); 
MOST and NSFC (China); 
CNRS/IN2P3 and CEA (France);  % added CEA 26/02/2026
BMFTR, DFG and MPG (Germany);
NKFIH (Hungary);              % added 16/06/2026
INFN (Italy); 
NWO (Netherlands); 
MNiSW and NCN (Poland); 
MEC/IFA (Romania); 
%MSHE (Russia); 
MICIU and AEI (Spain);
SNSF and SER (Switzerland); 
NASU (Ukraine); 
STFC (United Kingdom); 
DOE NP and NSF (USA).
%%%%%%%%%%%%%%%%%%%%%%%%%%%%%%%%%%%%%%%%%%%%%
We acknowledge the computing resources that are provided by ARDC (Australia), 
CBPF (Brazil),
CERN, 
IHEP and LZU (China),
IN2P3 (France), 
KIT and DESY (Germany), 
INFN (Italy), 
SURF (Netherlands),
Polish WLCG (Poland),
IFIN-HH (Romania), % http://dx.doi.org/10.13039/100019931,"Institutul National de Cercetare-Dezvoltare pentru Fizica si Inginerie Nucleara 'Horia Hulubei'"
%RRCKI and Yandex LLC (Russia), 
PIC (Spain), CSCS (Switzerland), 
GridPP (United Kingdom),
and NSF (USA).  % added Feb2026
%%%%%%%%%%%%%%%%%%%%%%%%%%%%%%%%%%%%%%%%%% 
We are indebted to the communities behind the multiple open-source
software packages on which we depend.
%%%%%%%%%%%%%%%%%%%%%%%%%%%%%%%%%%%%%%%%%%
Individual groups or members have received support from
% ARC and ARDC (Australia); % moved to national 16/01/2025
RTP (Australia), % added 06/03/2026
FWO Odysseus grant G0ASD25N (Belgium), % added 20/4/2026
Key Research Program of Frontier Sciences of CAS, CAS PIFI, CAS CCEPP (China); 
%Fundamental Research Funds for the Central Universities,  and Sci.\
%\& Tech.\ Program of Guangzhou (China); Removed 24/11/25
Minciencias (Colombia);
EPLANET, Marie Sk\l{}odowska-Curie Actions, ERC and NextGenerationEU (European Union);
A*MIDEX, ANR, IPhU and Labex P2IO, and R\'{e}gion Auvergne-Rh\^{o}ne-Alpes (France);
%RFBR, RSF and Yandex LLC (Russia);
Alexander-von-Humboldt Foundation (Germany);
ICSC (Italy); 
%GVA, XuntaGal, GENCAT, Inditex, InTalent and Prog.~Atracci\'on Talento, CM (Spain);
Severo Ochoa and Mar\'ia de Maeztu Units of Excellence, GVA, XuntaGal, GENCAT, InTalent-Inditex and Prog.~Atracci\'on Talento CM (Spain);
%XuntaGal --> Xunta de Galicia 
% SRC (Sweden);  % removed 27/02/2026 - end of grant
the Leverhulme Trust, the Royal Society and UKRI (United Kingdom).

%\input{LHCbInternal/supplementary_template}

%\input{appendix}

% This should be taken out in the final paper
%%\input{supplementary}

% This should be taken out in the final paper
%\input{LHCbInternal/data_and_software_availability_template}
\clearpage
\addcontentsline{toc}{section}{References}
%\setboolean{inbibliography}{true}
\bibliographystyle{LHCb/LHCb}
\bibliography{main,LHCb/standard,LHCb/LHCb-PAPER,LHCb/LHCb-CONF,LHCb/LHCb-DP,LHCb/LHCb-TDR,LHCb/LHCb-PUB}

\newpage
% LHCb collaboration author list
% Data extracted on September 8th, 2026 at 1:57am for paper reference LHCb-PAPER-2026-034
\centerline
{\large\bf LHCb collaboration}
\begin
{flushleft}
\small
R.~Aaij$^{40}$\lhcborcid{0000-0003-0533-1952},
M.~Abdelfatah$^{72}$,
A.S.W.~Abdelmotteleb$^{60}$\lhcborcid{0000-0001-7905-0542},
C.~Abellan~Beteta$^{54}$\lhcborcid{0009-0009-0869-6798},
F.~Abudin\'en$^{62}$\lhcborcid{0000-0002-6737-3528},
T.~Ackernley$^{64}$\lhcborcid{0000-0002-5951-3498},
A.A.~Adefisoye$^{72}$\lhcborcid{0000-0003-2448-1550},
B.~Adeva$^{50}$\lhcborcid{0000-0001-9756-3712},
M.~Adinolfi$^{58}$\lhcborcid{0000-0002-1326-1264},
P.~Adlarson$^{88,45}$\lhcborcid{0000-0001-6280-3851}
% ...

{\footnotesize \it

% ...
$^{40}$Nikhef National Institute for Subatomic Physics, Amsterdam, Netherlands\\
% ...
$^{45}$National Center for Nuclear Research (NCBJ), Warsaw, Poland\\
% ...
$^{50}$Instituto Galego de F{\'\i}sica de Altas Enerx{\'\i}as (IGFAE), Universidade de Santiago de Compostela, Santiago de Compostela, Spain\\
% ...
$^{54}$Physik-Institut, Universit{\"a}t Z{\"u}rich, Z{\"u}rich, Switzerland\\
% ...
$^{58}$H.H. Wills Physics Laboratory, University of Bristol, Bristol, United Kingdom\\
% ...
$^{60}$Department of Physics, University of Warwick, Coventry, United Kingdom\\
% ...
$^{62}$School of Physics and Astronomy, University of Edinburgh, Edinburgh, United Kingdom\\
$^{63}$School of Physics and Astronomy, University of Glasgow, Glasgow, United Kingdom\\
$^{64}$Oliver Lodge Laboratory, University of Liverpool, Liverpool, United Kingdom\\
% ...
$^{72}$Syracuse University, Syracuse, NY, United States\\
% ...
$^{88}$Department of Physics and Astronomy, Uppsala University, Uppsala, Sweden, associated to $^{63}$\\
% ...
}
\end{flushleft}

\end{document}